\documentclass{aastex63}

\submitjournal{The Astrophysical Journal}

\shorttitle{NGVS Distant RR Lyrae}
\shortauthors{Feng et al.}
\graphicspath{{./}{figures/}}

\begin{document}

\title{The Next Generation Virgo Cluster Survey. XXXVII.\\
Distant RR Lyrae Stars and the Milky Way Stellar Halo out to 300~kpc}

\correspondingauthor{Yuting Feng}
\email{yfeng47@ucsc.edu}

\author{Yuting Feng}
\affiliation{Department of Astronomy and Astrophysics, University of California Santa Cruz, 1156 High Street, Santa Cruz, CA 95064, USA}

\author[0000-0001-8867-4234]{Puragra Guhathakurta}
\affiliation{Department of Astronomy and Astrophysics, University of California Santa Cruz, 1156 High Street, Santa Cruz, CA 95064, USA}

\author[0000-0002-2073-2781]{Eric W. Peng}
\affiliation{National Optical-Infrared Astronomy Research Laboratory (NOIRLab), 950 North Cherry Avenue, Tucson, AZ 85719, USA}

\author[0000-0001-8221-8406]{Stephen D. J. Gwyn}
\affiliation{NRC Herzberg Astronomy and Astrophysics, 5071 West Saanich Road, Victoria, BC, V9E 2E7, Canada}

\author[0000-0002-8224-1128]{Laura Ferrarese}
\affiliation{NRC Herzberg Astronomy and Astrophysics, 5071 West Saanich Road, Victoria, BC, V9E 2E7, Canada}

\author[0000-0003-1184-8114]{Patrick Côté}
\affiliation{NRC Herzberg Astronomy and Astrophysics, 5071 West Saanich Road, Victoria, BC, V9E 2E7, Canada}

\author[0000-0002-3263-8645]{Jean-Charles Cuillandre}
\affiliation{AIM, CEA, CNRS, Université Paris-Saclay, Université de Paris, F-91191 Gif-sur-Yvette, France}

\author{Jeffrey Munsell}
\affiliation{Bronx High School of Science, Bronx, NY 10468, USA}
\affiliation{Cornell University, Ithaca, NY 14850, USA}

\author{Manjima Talukdar}
\affiliation{St. Xavier’s College, 30 Park Street,
Kolkata 700016, India}
\affiliation{Department of Physics and Astronomy, Clemson University, 118 Kinard Laboratory, Clemson, SC 29634, USA}






\begin{abstract}
\noindent
RR Lyrae stars are standard candles with characteristic photometric variability, and serve as powerful tracers of Galactic structure, substructure, accretion history, and dark matter content. Here we report the discovery of distant RR Lyrae stars, including some of the most distant stars known in the Milky Way halo, with Galactocentric distances of $\sim300$~kpc. We use time-series $u^*g’i’z’$ Canada-France-Hawaii Telescope/MegaCam photometry from the Next Generation Virgo Cluster Survey (NGVS). We use a template light curve fitting method based on empirical Sloan Digital Sky Survey (SDSS) Stripe 82 RR~Lyrae data to identify RR~Lyrae candidates in the NGVS data set. We eliminate several hundred suspected quasars, and identify 180 RR Lyrae candidates, with heliocentric distances of $\sim20$--300~kpc. The halo stellar density distribution is consistent with an $r^{-4.09\pm0.10}$ power-law radial profile over most of this distance range with no signs of a break. The distribution of ab-type RR Lyrae in a period-amplitude plot (Bailey diagram) suggests that the mean metallicity of the halo decreases outwards. Compared to other recent RR Lyrae surveys, like Pan-STARRS1 (PS1), the High Cadence Transient Survey (HiTS), and the Dark Energy Survey (DES), our NGVS study has better single-epoch photometric precision and a comparable number of epochs, but smaller sky coverage. At large distances, our RR Lyrae sample appears to be relatively pure and complete, with well measured  periods and amplitudes. These newly discovered distant RR~Lyrae stars are important additions to the few secure stellar tracers beyond 150~kpc in the Milky Way halo.
\end{abstract}



\section{Introduction} \label{sec:intro}
%
%
The stellar halo of the Milky Way (MW) preserves much of the archaeological evidence of the Galaxy's accretion and assembly history \citep{2005ApJ...635..931B}. Many major discoveries have been made from recent surveys of the stellar halo. The ‘Gaia sausage’ or ‘Gaia-Enceladus’ substructure discovered in the 3D velocity space of main sequence stars \citep{2018MNRAS.478..611B, 2018Natur.563...85H} suggests a significant merger event of the MW with a dwarf galaxy. Results from the H3 spectroscopic survey of red giant stars in conjunction with Gaia proper motion measurements \citep{2020ApJ...901...48N} indicate that the stellar halo within 50~kpc is entirely comprised of substructure. The HALO7D survey \citep{2019H7Da, 2019H7Db} and its precursor survey of the MW halo in the foreground of the Andromeda galaxy \citep{13M31foregrounda, 16M31foregroundb} found evidence of substructure in the 3D kinematics of main sequence stars. We are in the midst of a Galactic Renaissance, and our perception of the disk, inner halo (Galactocentric radii $R_{\rm GC} \lesssim 50$~kpc), and disk-halo interface of our MW galaxy is being revolutionized.

The outer region of the MW halo ($R_{\rm GC} \gtrsim 50$~kpc) is an excellent test bed to study the assembly history of our Galaxy. This region provides unique leverage for measuring the overall extent and total mass of our Galaxy, which are key parameters for Galactic studies and near-field cosmology. Models suggest that stars in the outer halo ($R_{\rm GC}>100$~kpc) likely originated in recently accreted satellite galaxies \citep[e.g.,][]{2005ApJ...635..931B,2009ApJ...702.1058Z}. Current models of galaxy formation generate specific, but not yet observationally proven, predictions of the ``splash-back'' radius of the outermost part of the Galactic halo, which is physically defined as the radius where gravitationally captured particles reach the apocenter of their first orbit \citep{2020MNRAS.496.3929D, 2021MNRAS.504.4649O, 2021ApJ...915L..18L}. 
Exploring these issues greatly benefits from having a uniform sample of field star tracers with reliable distance estimates across the full radial extent of the Galactic halo, bridging the well-studied inner halo and the outer reaches mainly probed by dwarf satellites \citep[e.g.,][]{2019MNRAS.484.5453C}.

Only a small number of remote MW halo field stars have been detected at Galactocentric distances ($R_{\rm GC}$) larger than 100~kpc. \citet{2014ApJ...790L...5B} reported the discovery of two distant M giants with estimated distances larger than 200 kpc. These stars are intrinsically bright, making them good tracers of halo structure at relatively large distances, but their distance estimates suffer from significant uncertainties \citep[$\sim25$\% distance uncertainty according to][]{2014ApJ...790L...5B}. This, combined with an overall low identification accuracy \citep[$\sim20$\% according to][]{2014ApJ...790L...5B} make these RGB stars less ideal as tracers of the outer stellar halo.

RR Lyrae variable stars are old ($\rm age > 10~Gyr$) horizontal branch stars that pulsate with short periods (0.2--1.2~d). They have become one of the most widely used stellar tracers in MW and Local Group studies. RR Lyrae stars were discovered more than a century ago \citep{1901ApJ....13..226P} and there have been extensive studies of the nature and characteristics of the pulsation in these stars (see \cite{2015pust.book.....C} for a review). These stars have well-constrained period-luminosity-metallicity ($P$-$L$-$Z$) relations \citep[e.g.,][]{2004ApJS..154..633C, 2015ApJ...808...50M}, making them excellent distance indicators. This, combined with their bright luminosities ($M_V \sim +0.6$) and advanced ages, make RR Lyrae suitable for precisely tracing the stellar populations and substructures (satellite galaxies, star clusters, and streams) within the MW halo. Recent simulation work \citep[e.g.,][]{2017MNRAS.470.5014S} suggest that there should be thousands of RR Lyrae stars in the MW halo beyond $R_{\rm GC}>100$ kpc, and they mostly originate from dwarf satellites recently accreted by our galaxy. We can further investigate these RR Lyrae variables' photometric metallicities using their light curve shapes \citep{2013ApJ...773..181N}, or even infer the elemental abundance patterns of their formation environments based on their distribution in the period-amplitude diagram \citep[also known as a Bailey diagram, see][]{2009Ap&SS.320..261C, 2018MNRAS.477.1472B}. Once these distant RR Lyrae populations are identified, their distribution and pulsational properties provide us valuable clues about the formation history, radial density profile, and mass of the MW halo. 

RR Lyrae stars have been useful probes of the inner stellar halo for decades \citep[e.g.,][]{1966ApJS...13..379K,1984ApJ...283..580S,1984MNRAS.206..433H,1985ApJ...289..310S,1989AJ.....98.1648C}, but the advent of wide-field, time-domain imaging on a variety of telescopes has allowed for more comprehensive surveys of the outer halo. The efficacy and precision of light curve template fitting for determining the pulsational parameters of RR Lyrae stars have been well-established since the technique's initial applications in the last century \citep{1996PASP..108..877J, 1998AJ....115..193L}. Several recent studies have used both template-fitting algorithms and visual inspections to identify increasingly distant RR Lyrae stars in the MW halo, all with survey data sets that have sparse and non-uniform temporal coverage. \citet{2017AJ....153..204S} applied lightcurve template fitting techniques on stars observed in the Pan-STARRS1 (PS1) 3$\pi$ data, and reported the discovery of over 45,000 RRab stars (the fundamental-mode subtype of RR Lyrae) out to $\sim 130$ kpc. \citet{2018ApJ...855...43M} pushed the distance limit of RR Lyrae detection to $\sim 250$ kpc with the High Cadence Transient Survey (HiTS) dataset. Most recently, \citet{2021ApJ...911..109S} presented their catalog of 6,971 newly discovered RRab stars, with the most distant candidate at $R_{\rm GC} \sim 330$ kpc, based on their analysis of the six-year Dark Energy Survey (DES Y6) data. However the single-epoch photometric precision of all three surveys is not particularly high beyond $g>22$. This, combined with limited survey cadence, gets in the way of robust identification of RR Lyrae stars beyond $R_{\rm GC}>200$~kpc. 

In our work, we utilize the Next Generation Virgo Cluster Survey \citep[NGVS;][]{2012ApJS..200....4F}, which has significantly higher single-epoch photometric precision than PS1, HiTS, and DES, and a comparable number of epochs. This allows us to robustly identify a sample of distant RR Lyrae sample in the MW halo. 

This paper is structured as follows. In Section \ref{sec:data}, we describe the NGVS time-domain photometry, as well as our variable object selection criteria. In Section \ref{sec:method}, we describe the process of identification and characterization of the NGVS RR Lyrae sample. In Section \ref{sec:results}, we present and discuss the analysis of this sample, including the spatial distribution, density profile and the Bailey diagram. In Section \ref{sec:future}, we discuss future work and summarize our main results.


\begin{deluxetable}{cccccc}
\tablenum{1}
\tablecaption{Basic Parameters of NGVS Observations\label{tab:NGVS_params}}
\tablewidth{0pt}
\tablehead{ 
\colhead{Band} & \colhead{Epochs}  & \colhead{Exp.\ Time} & \multicolumn{2}{c}{Detection Limit} &  \colhead{FWHM} \\
 & & & stacked & single & \\
\colhead{} &  & \colhead{[s]} & \colhead{[mag]} & \colhead{[mag]} & \colhead{[arcsec]} \\
\colhead{(1)} &
\colhead{(2)} &
\colhead{(3)} &
\colhead{(4)} &
\colhead{(5)} &
\colhead{(6)} 
}
\startdata
$u^*$ & 13 & 582 & 26.3 & 24.5 & $\leq 1\farcs1$\\
$g'$ & 7 & 634 & 25.9 & 24.4 & $\leq 1\farcs0$\\
$i'$ & 8 & 411 & 25.1 & 23.6 & $\leq 0\farcs6$\\
$z'$ & 10 & 550 & 24.8 & 23.1 & $\leq 1\farcs0$
\enddata
\tablecomments{\\
(1) CFHT MegaCam filter/band used. \\
(2) Median number of epochs in each band.\\
(3) Exposure time in seconds [s].\\
(4) and (5) Point source detection limits in the stacked and individual exposures, respectively. The detection limits are $10\sigma$ for the $g'$ and $i'$ bands, and $5\sigma$ for the $u^*$ and $z'$ bands.\\
(6) Image quality full width at half maximum (FWHM) criteria for NGVS queue observations.
}
\end{deluxetable}

\section{Time Series Photometry in the NGVS}\label{sec:data}
\subsection{General Survey Information}\label{subsec:data_general_info}
The Next Generation Virgo Cluster Survey (NGVS) is a deep optical imaging survey of the Virgo cluster of galaxies, carried out with the 1~deg$^2$ MegaCam instrument on the 3.6-m Canada-France-Hawaii Telescope (CFHT) in the $u^*$, $g'$, $i'$, and $z'$ bands (with limited additional coverage in the $r'$ band---not used in this work). The survey consists of 117~pointings (not including 4~background pointings), with slight overlap between adjacent pointings. The resulting NGVS footprint covers a contiguous area of 104 deg$^2$ of the nearby Virgo cluster, out to the virial radii for both the Virgo~A and B subclusters. The NGVS project motivations, strategy, and observational program are discussed by \citet{2012ApJS..200....4F}. The main goal of the NGVS is to study the galaxy population of the Virgo cluster. The observing strategy adopted, however, was atypical of deep, extragalactic optical surveys, and involved imaging multiple fields in sequence before returning to the same field \citep[see \S\,2 of][for a full description]{2012ApJS..200....4F}. The purpose of this strategy was to improve sky subtraction on large scales, and it naturally resulted in an observing cadence where individual observations of a single field were spaced by a minimum of $\sim\!1$~hr and by as much as a few yr. This range of temporal spacing makes the NGVS data well-suited to finding both short- and long-term variables.

The NGVS images reach point source depths of [$u^*$ ($S/N=5$), $g'$ ($S/N=10$), $i'$ ($S/N=10$), $z'$ ($S/N=5$)]$\;\approx\;$[26.3, 25.9, 25.1, 24.8] mag for the stacked images, and [24.5, 24.4, 23.6, 23.1]~mag for single exposures. This paper is based on all NGVS single exposure images obtained during the period March 2009 through May 2013.  The total number of epochs across the four bands is roughly the same ($\sim\!38$) for each MegaCam pointing (tile in the survey mosaic), with the exception of small overlap regions between adjacent pointings which have twice as many epochs or more. The single-exposure integration times [and approximate number of exposures per band] are 582$\;$s [$\times13$] in $u^*$,
634$\;$s [$\times7$] in $g'$,
411$\;$s [$\times8$] in $i'$, and
550$\;$s [$\times10$] in $z'$.
The sky was clear during most exposures, with a representative seeing FWHM $\sim0\farcs8$, although the image quality in $i'$ was always better than $0\farcs6$. The basic parameters of the NGVS dataset are summarized in Table\,\ref{tab:NGVS_params}. The NGVS dataset represents the deepest uniform imaging available across the entire Virgo cluster, and is likely to remain so for some time. 

\begin{figure}
    \centering
    \includegraphics[width=0.5\textwidth]{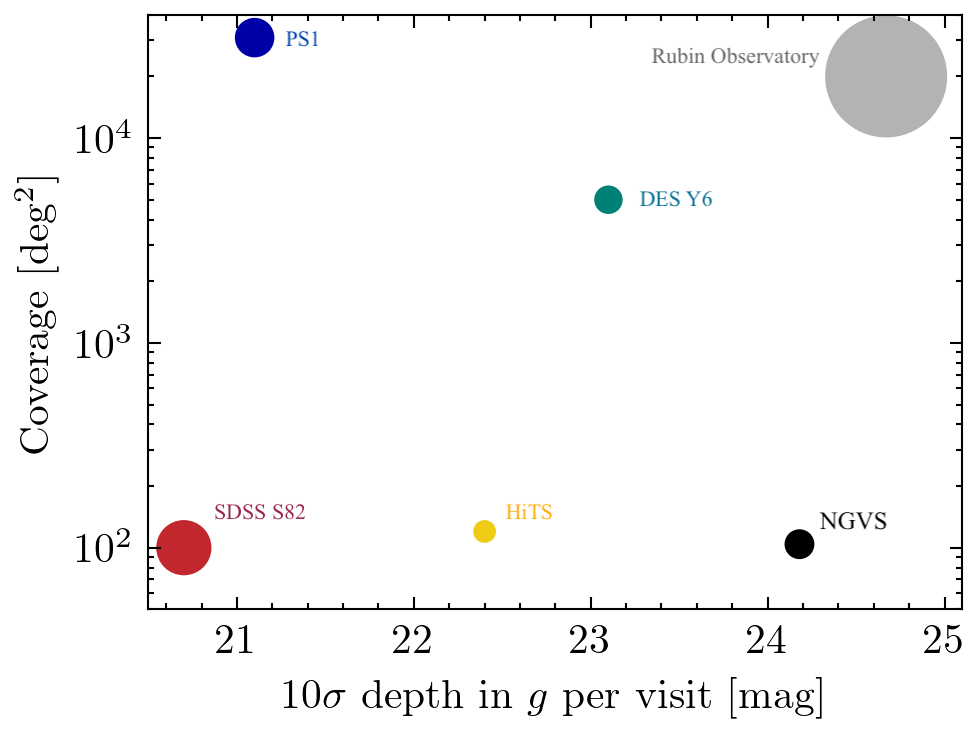}
    \caption{A comparison of NGVS to other recent and upcoming time-domain surveys of RR Lyrae in the outer halo of the MW. The radius of each data point is proportional to the total number of epochs across all bands. The area surveyed by the NGVS is comparable to SDSS Stripe 82 and HiTS, and its single-epoch photometric depth is $\sim\!1$~mag fainter than wider surveys like DES Y6. For reference, Rubin/LSST is also plotted to show how it will equal or surpass all surveys preceding it along all three axes of this parameter space.}
    \label{fig:NGVS_compare}
\end{figure}

Figure~\ref{fig:NGVS_compare} shows how the NGVS compares to recent and future surveys of RR Lyrae stars in terms of survey area, number of epochs, and depth. The NGVS occupies an interesting part of this survey parameter space: its area is comparable to Sloan Digital Sky Survey (SDSS) Stripe 82 and HiTS, its number of epochs is comparable to PS1 and DES Y6, and its single-epoch depth is $\sim\!1$~mag fainter than DES. Only with Rubin/LSST will there be a survey of the distant MW halo with better single epoch depth than NGVS and with a substantially larger area.

\subsection{Photometry and Point Source Selection}
\label{subsec:stacked_phot}
The pre-processed individual NGVS images were stacked for each of the four bands, and the stacked images in the four bands were re-gridded onto a common astrometric system. These steps were carried out using the {\tt MegaPipe} software \citep[see][]{2008PASP..120..212G, 2012ApJS..200....4F}. The {\tt SExtractor} software was used in ``dual image mode'' to detect sources in the $g'$ band and carry out forced photometry in all four bands  (of the four, $g'$ is the one in which faint RR~Lyrae have the highest signal-to-noise ratio in the stacked NGVS images). Aperture photometry was carried out for all detected objects on the stacked NGVS $u^*$, $g'$, $i'$, and $z'$ images using the following set of aperture diameters: 3, 4, 5, 6, 7, 8, and 16 pixels, and a local sky annulus of $(r_{\text{inner}},~ r_{\text{outer}}) = (16,~ 20)$~pixels. where each pixel corresponds to 0\farcs187.

A three-step calibration/correction process was applied in order to obtain optimal point-source photometry, while accounting for the fact that the point spread function (PSF) varies from field to field across the NGVS footprint and from band to band. First, a sample of bright (but unsaturated in NGVS) stars were selected in each field and their 16-pixel (3\farcs0) aperture instrumental magnitudes were photometrically calibrated to their SDSS PSF magnitudes. Second, a map of aperture correction as a function of sky position was created based on the curves of growth of a larger sample of bright stars (extending to fainter magnitudes than the SDSS calibration stars) in each of the four bands. Third, the aperture magnitudes (diameters of 3, 4, 5, 6, 7, and 8 pixels) of all objects were corrected to 16-pixel aperture total magnitudes based on the aperture correction map. The 16-pixel aperture contains practically all of the flux of a point source, given the sub-arcsecond to arcsecond (FWHM) seeing conditions under which the NGVS data were obtained (Table~\ref{tab:NGVS_params}). In this way, we have compiled a homogeneous point-source photometry database that is calibrated against SDSS. To summarize, $m_{b,N}$ for each point source represents its total magnitude in band $b$, aperture-corrected from its $N$-pixel aperture magnitude on the stacked image. In this study, we use the aperture-corrected 4-pixel ($0\farcs75$) diameter total apparent magnitudes in the four bands, $u^*_4$, $g'_4$, $i'_4$, and $z'_4$ ; they are hereafter referred to simply as $u^*$, $g'$, $i'$, and $z'$ magnitudes, respectively.\footnote{The stacked-image photometry used through most of this paper is the flux-averaged apparent brightness over the NGVS cadence. This cadence is {\bf sparse} and varies from band to band for any given source. For our NGVS RR~Lyrae candidates, we compute the {\bf complete} flux-averaged apparent magnitude in each band $b$, $\langle{m_b}\rangle$, by integrating over the full light curve based on the best-fit pulsational parameters (see \S\S\,\ref{sec:method}--\ref{sec:future}).}

\begin{figure}
    \centering
    \includegraphics[width=0.5\textwidth]{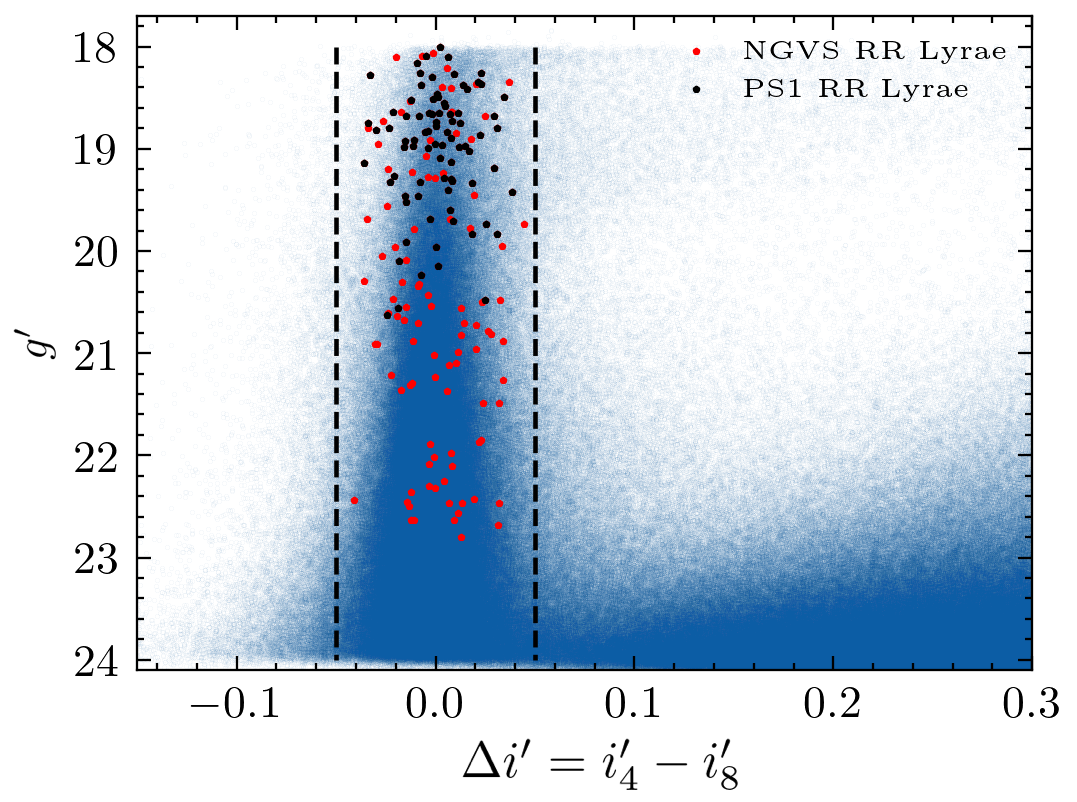}
    \caption{Distribution of the $i'$ band fuzziness index $\Delta i' = i'_4 - i'_8$ versus the $g'$ apparent magnitude of all sources extracted from the NGVS stacked images (small blue dots). The vertical sequence of point sources at $\Delta i'\sim0$ is clearly visible. The recovered PS1 RR~Lyrae in this study are marked as filled black dots, while the newly identified NGVS RR~Lyrae candidates are plotted as filled red dots. The black dashed lines at $\Delta i' =\pm 0.05$ shows our selection range for point sources.}
    \label{fig:fuzziness}
\end{figure}

At the faint apparent magnitudes that this study explores, any survey of stars must contend with contamination by background galaxies and quasi-stellar objects (QSOs). In order to distinguish between point sources and extended sources, we adopt a fuzziness index: $\Delta i' = i'_4 - i'_8$. An object is considered to be a point source if $-0.05\leq \Delta i' \leq +0.05$. With the above definition of $m_{b,N}$, we expect the distribution of $m_{b, N_1} - m_{b, N_2}$ for point sources to concentrate around 0. Of the four NGVS bands, we chose to use $i'$ magnitudes because: (1)~the $i'$ images were obtained under the best seeing conditions ($\rm FWHM\sim0$\farcs6); and (2)~the field-to-field variance in the seeing FWHM was smallest in the $i'$ band. Our point source selection criteria are similar to those used in previous NGVS studies \citep{2014ApJ...794..103D, 2015ApJ...812...34L}. The distribution of $\Delta i'$ versus $g'$ magnitude shown in Figure~\ref{fig:fuzziness} displays a clear vertical locus of point sources (mostly stars, along with some QSOs) around $\Delta i' = 0$. The rest of this paper is limited to NGVS point sources as defined by the above $\Delta i'$ criteria.

\subsection{Color-Color Selection Based on Known RR Lyrae} \label{sec:clr_clr_sel}
\begin{figure*}
\gridline{\fig{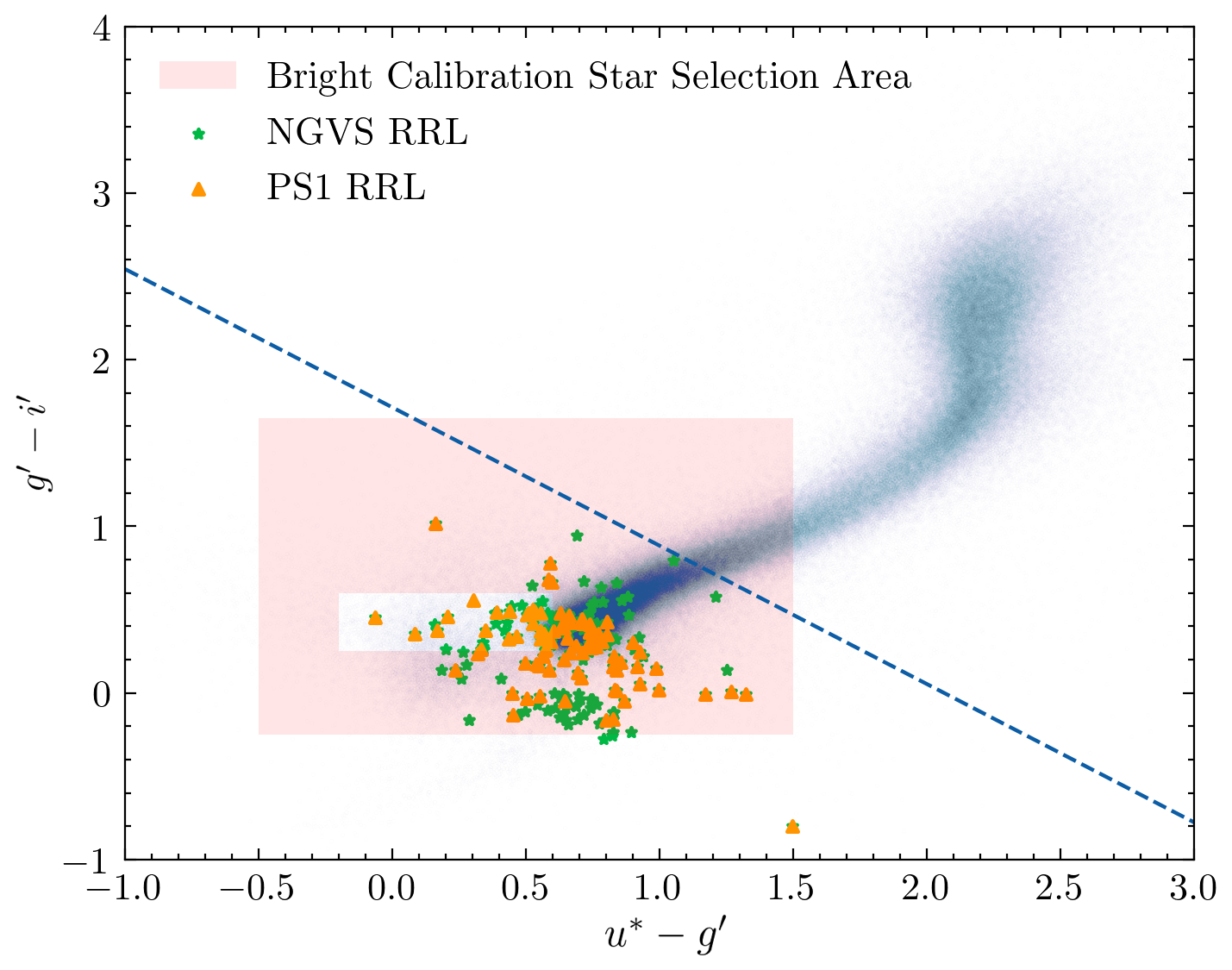}{0.45\textwidth}{(a)}
\fig{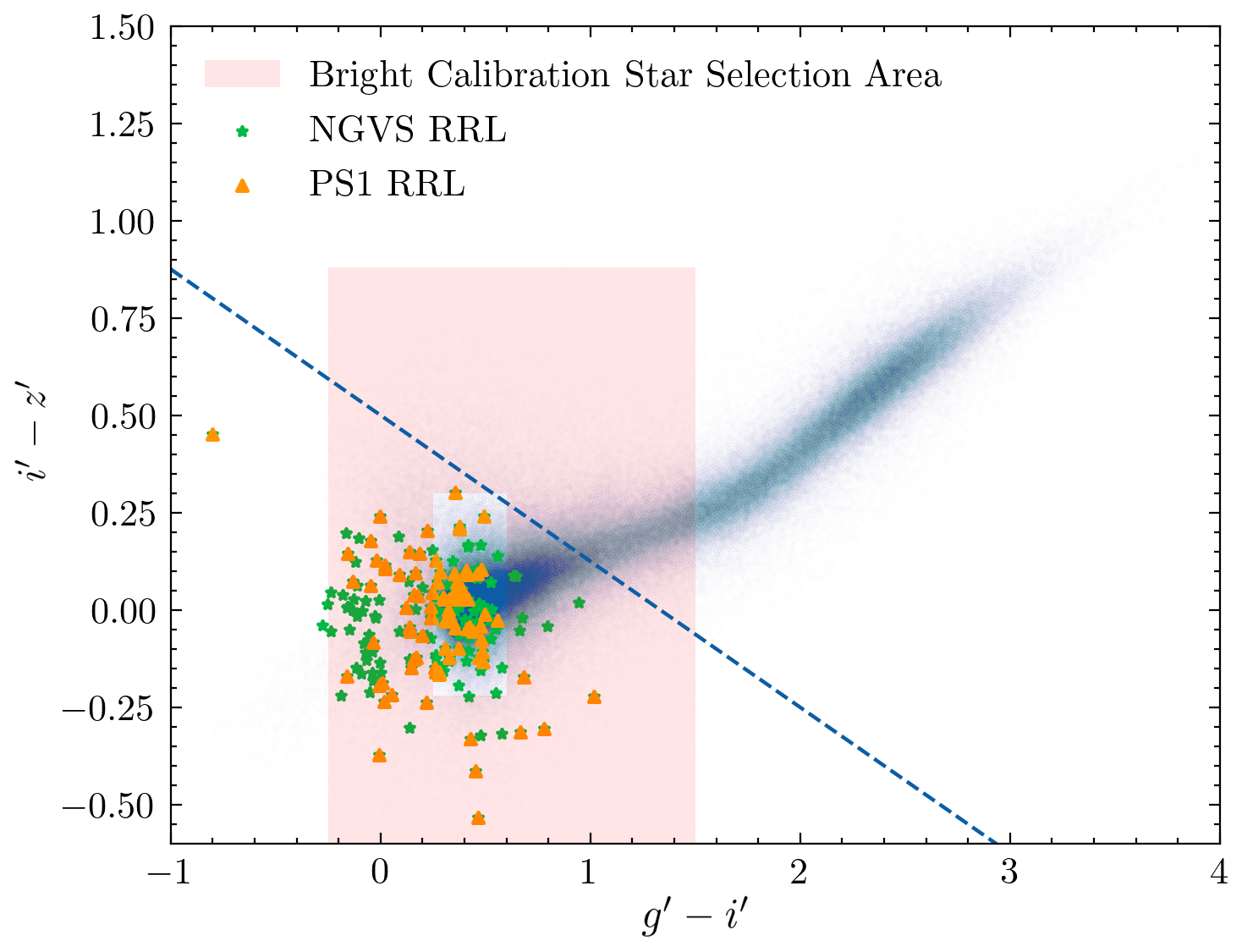}{0.45\textwidth}{(b)}
          }
\caption{Color-color diagrams, $g'-i'$ versus $u^*-g'$ ({\it left\/}) and $i'-z'$ versus $g'-i'$ ({\it right\/}), for all point sources in the NGVS (small blue dots). The dashed blue diagonal lines demarcate the sections of these two color-color diagrams that are occupied by known PS1 RR~Lyrae. In our search for NGVS RR~Lyrae, all point sources that are located to the lower left of the lines in both color-color diagrams were searched for variability. PS1 RR~Lyrae candidates in the NGVS footprint that were recovered in our search and newly discovered NGVS RR~Lyrae candidates are marked as orange triangles and green stars, respectively. Stars that are located in the pink shaded region of both color-color diagrams and satisfy $g'<21$ are selected as photometric reference stars (see \S\,\ref{subsec:ap_phot_calib}). The rectangular unshaded region within the pink shaded rectangular region in each of the two color-color diagrams excludes some obvious bright RR~Lyrae.
}
\label{fig:clrclr}
\end{figure*}

To search for RR~Lyrae in the NGVS database, we first select all point sources in the region of $(u^*-g')$ versus $(g'-i')$ and $(g'-i')$ versus $(i'-z')$ color-color space that is occupied by known RR~Lyrae---i.e., those that are located to the lower left of the blue dashed lines in both color-color diagrams (Figure~\ref{fig:clrclr}). Since these colors are based on stacked-image photometry, they can be quite different from the {\it true\/} colors for variable objects, because the NGVS cadence is sparse and different from band to band. In other words, it is not surprising that known RR~Lyrae display a relatively large scatter in our color-color diagrams. The dashed lines we use take this scatter into account.

Of the 94 RR~Lyrae discovered by \cite{2017AJ....153..204S} in the NGVS footprint using  PS1 survey data: 84 are confirmed as RR Lyrae in this work (see \S\,\ref{sec:method}); 7 are bright stars ($g'\sim18.0$--18.5) for which a portion of the NGVS time-series photometry in one or more of the bands suffers from saturation; and 3 are fainter objects ($g\sim20.5$) that are classified as QSOs or non-variable in our work. Our final color-color selection criteria, shown as blue dashed lines in Figure~\ref{fig:clrclr}, were designed to include all 84 of these reconfirmed PS1 RR~Lyrae whose photometry is free of saturation in NGVS.

We also defined regions around the stellar locus in the two color-color diagrams, as indicated by the pink shaded regions in Figure~\ref{fig:clrclr}, to select bright ($18<g'<21$), non-variable stars. This yields 
54,043~point sources ($\sim520$ per NGVS pointing) that are used as photometric reference stars for the calibration of our time-series photometry, as described in \S\,\ref{subsec:ap_phot_calib} below. Of these, 402 sources (0.7\%) turned out to be variable candidates (see \S\,\ref{subsec:variable}); the variable fraction is so low that it has a negligible effect on the calibration of the NGVS time-series photometry.

\subsection{Time Series Aperture Photometry and Calibration}\label{subsec:ap_phot_calib}
Our study is the first to explore time-domain photometry in the NGVS database. As a result, we had to develop specific data analysis methods to account for exposure-to-exposure temporal variations in the seeing FWHM in each band and residuals in the atmospheric transparency correction.

Each NGVS exposure is calibrated using astrometric solutions from the {\tt MegaPipe} image processing pipeline \citep{2008PASP..120..212G,2012ApJS..200....4F}. Aperture photometry was carried out for all point sources using the {\tt photutils} Python package \citep{larry_bradley_2020_4044744} on small $12\farcs15 \times 12\farcs15$ ``postage stamps'' which were cut out from single epoch NGVS images around our point sources of interest. The apparent magnitude of the central point source in each cutout image is measured using an aperture whose radius is equal to the best-fit average FWHM of the PSF for that exposure, as determined by {\tt MegaPipe} before image stacking. Aperture radii range from 0\farcs56 to 1\farcs12 for the postage stamps analyzed in this paper. Scaling the aperture size to the PSF FWHM optimizes the photometric signal-to-noise. For the purpose of sky subtraction, the background in each postage stamp is defined to be the median brightness within an annulus of $(r_{\text{inner}},~ r_{\text{outer}}) = (2\farcs62,~ 3\farcs37)$. For each photometric reference star on each exposure, we measure the aperture correction:
\begin{equation}
    \Delta (m_b)^{\text{exp}} = (m_b)_{\text{aper}}^{\text{exp}} - m_b
\end{equation}
\noindent
for all photometric reference stars on a given exposure, where $(m_b)_{\text{aper}}^{\text{exp}}$ is the aperture magnitude of the reference star on a given exposure in band $b$. We then calculate the median value of the aperture correction, $\Delta (m_b)^{\text{exp}}$, for all photometric reference stars on the exposure. This median aperture correction is applied to the FWHM-optimized aperture magnitudes of all point sources of interest on that exposure. This photometric procedure accounts for variations in atmospheric transparency and PSF quality across the different exposures.

\begin{figure}
    \centering
    \includegraphics[width=0.45\textwidth]{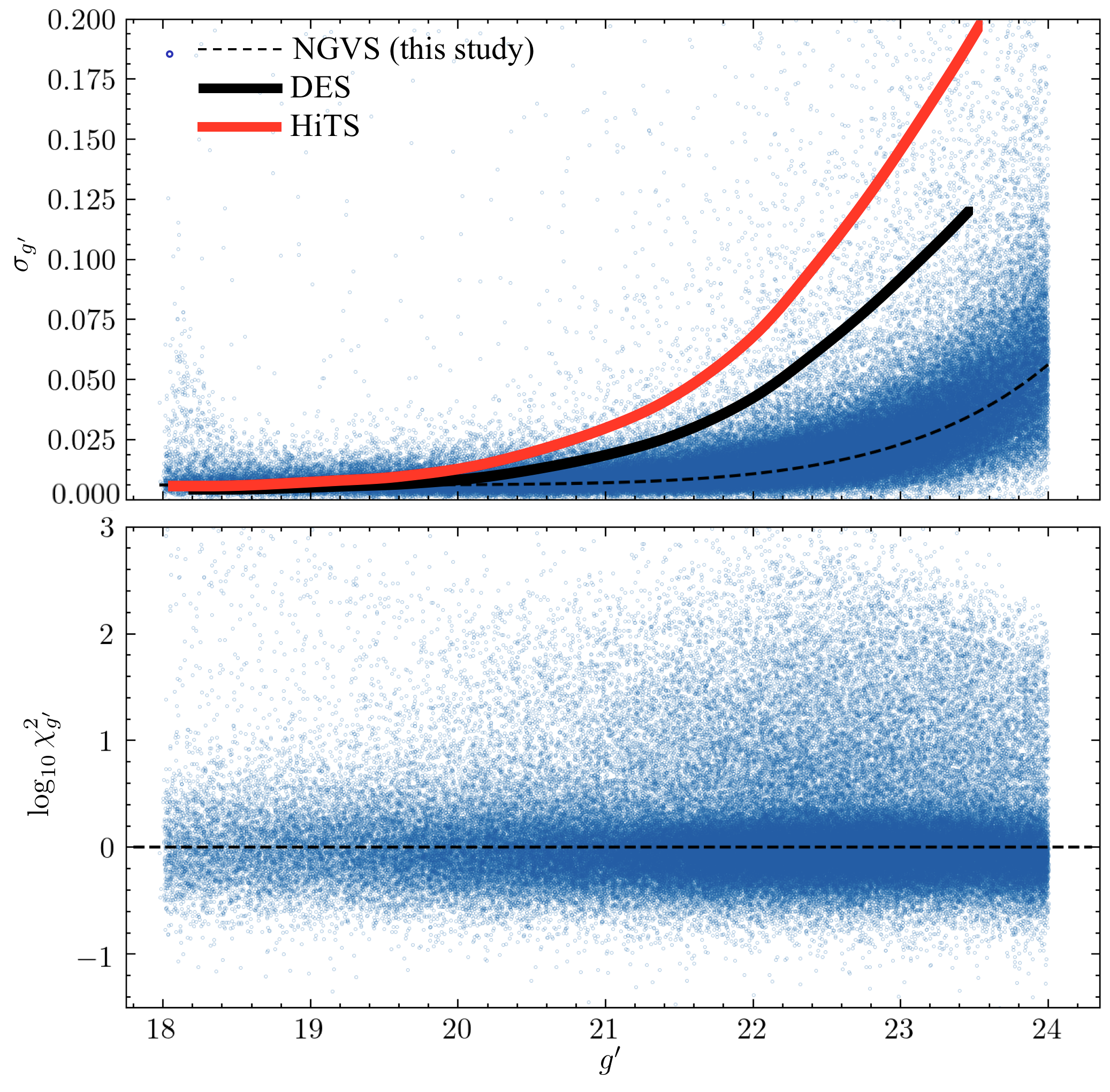}
    \caption{{\it Upper panel\/}: The distribution of the empirical RMS dispersion $\sigma_{g'}$ versus apparent magnitude $g'$ for all point sources in NGVS. The dashed back line is a polynomial fit to the running median of the RMS as a function of $g'$; the bold solid back and red lines are the corresponding trend lines for DES and HiTS, respectively. {\it Lower panel\/}: Plot of $\log_{10}{\chi^2_{g'}}$ versus apparent $g'$ magnitude based on our estimate of the total photometric error (quadrature sum of random and systematic error).}
    \label{fig:chi_std}
\end{figure}

To estimate the systematic uncertainty of our photometry, and to ensure our measurements across different epochs are self-consistent, we did an error-rescaling by calibrating the reduced chi-squared statistic, $\chi_{\nu,\,b}^2$. This was calculated using the median magnitude $\overline{m_{b}}$ in a given band $b$ for each light curve:
\begin{equation}
     \chi_{\nu,\,b}^2 = \frac{1}{N_{b}-1}\sum_{1}^{N_b}\frac{(m_{i,\,b}-\overline{m_{b}})^2}{\sigma_{i,\,b}^2}
\end{equation}
in which $N_b$ is the number of epochs of star $\nu$ in band $b$, $m_{i,\,b}$ is the $i^{\rm th}$ measurement in that band, and $\sigma_{i,\,b}$ is the photometric uncertainty which is a combination of random (Poisson) error and systematic error:
\begin{equation}
    \sigma_{i,\,b} = \sqrt{\sigma_{i,\,b,\, \text{random}}^2 + \sigma_{b,\,\text{sys}}^2}
\end{equation} 
We assume that the systematic error $\sigma_{\rm sys}$ should consist of two terms: a constant term $\sigma_{0,\,b}$ that accounts for bright nearby objects and/or bad sky subtraction, and a linear term $\sigma_{1,\,b}\times10^{\frac{\langle m_{\nu,\,b} \rangle}{2.5}}$ that scales with the apparent brightness of the star and accounts for effects like CCD non-homogeneity. For non-variable sources, which are the majority of our sample, it is reasonable to expect that $\chi_{\nu,\,b}^2 \sim 1$. Therefore, the parameter values $(\sigma_{0,\,b},\,\sigma_{1,\,b})$ that quantify the systematic error in each band $b$ are optimized to ensure that the $\log\,\chi_{\nu,\,b}^2$ versus $m_{b}$ trend line is horizontal and centered on zero, as shown in the lower panel of Figure~\ref{fig:chi_std}. This parameter optimization involves minimizing the sum of sigma-clipped $\left| \log{\chi_{\nu,\,b}^2} \right |$ for all NGVS stars in each band $b$. The photometric uncertainties reach $\sim 0.05$ mag at $g'\sim24$ (upper panel of Figure~\ref{fig:chi_std}). Compared to the two predecessor surveys, our NGVS single-epoch photometry is 1.7~mag deeper than DES and 2.3~mag deeper than HiTS at an error level of 0.05~mag.

\subsection{Identification of Variable Objects}\label{subsec:variable}
After the time-series photometry is calibrated, we select variable candidates that satisfy all three of the following criteria: (1) $\sigma_{\nu,\,b} \geq \sigma_{\text{med},\,b}$; (2) $\ln{\chi^2_{\nu,\,b}}\geq2$; and (3) $N_{\nu,\,b}\geq3$, where the meanings of subscripts are the same as in \S\,\ref{subsec:ap_phot_calib}. In other words, if an object $\nu$ has elevated photometric RMS and $\chi^2$ in all four bands relative to the corresponding median values for objects of comparable apparent magnitude, and has at least three photometric measurements in each band, it is considered a variable candidate. In total, 1685 variable candidates passed the variability cuts and are used as the input for our template fitting algorithm.

\section{Identification and Characterization of the NGVS RR Lyrae Sample}\label{sec:method}
\begin{figure*}
\gridline{\fig{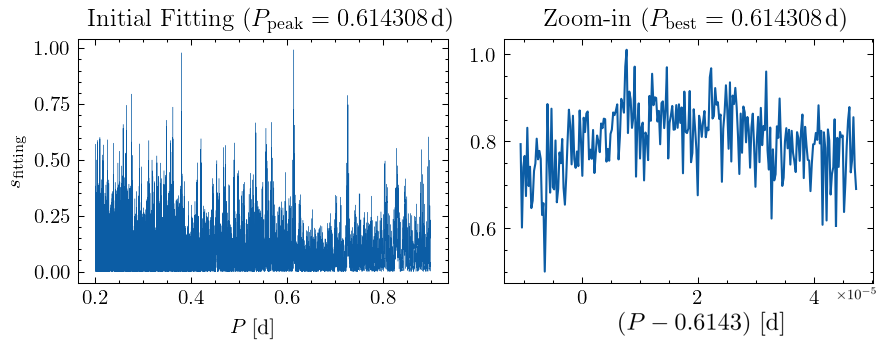}{0.85\textwidth}{(a)}}
\gridline{
\fig{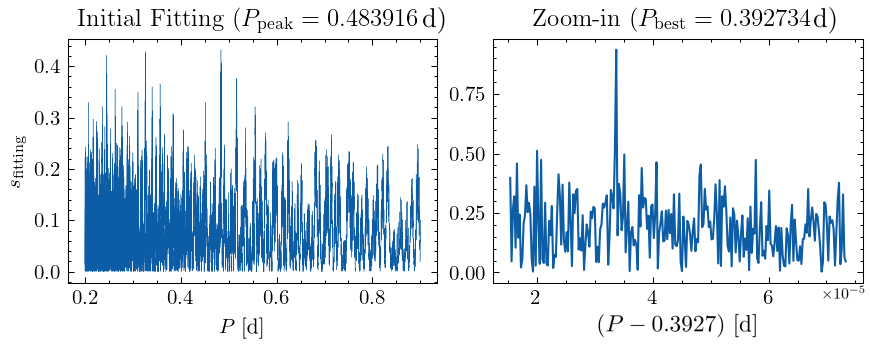}{0.85\textwidth}{(b)}
          }
\caption{Two periodogram ($P$ versus $s_{\rm fitting}$) cases of our template fitting of RR~Lyrae light curves. {\it Top\/}~(a): The left panel shows the initial fitting result, with $\Delta t = 0.5$~s. The tallest peak is at 0.614308~d. The right panel shows the zoom-in fitting result with $\Delta t = 0.02$~s around the tallest peak, while 0.614308~d remains the best-fit period. {\it Bottom\/}~(b): The left panel shows a tallest peak at 0.483916~d after the initial fitting, while the right panel shows the zoom-in fitting result around the 9th tallest peak of the initial fitting, which actually contains the global best-fit period of 0.392734~d. An extra expanded search with a smaller period interval around peaks obtained from the initial fitting is necessary for our fitting analysis of sparsely sampled RR Lyrae light curve.}%
    \label{fig:periodo}%
\end{figure*}

\subsection{Initial Light Curve Template Fitting} \label{subsec:initial}

To identify RR Lyrae and derive robust estimates of their light curve parameters like period and amplitude, we performed template fitting based on the empirical RR~Lyrae light curves generated from high cadence observations of 483 RR~Lyrae in the SDSS Stripe~82 \citep{2010ApJ...708..717S}. Of these 483~templates, 379 are of type RRab and 104 are of type RRc. Our light curve model has the form:
\begin{equation}
    m_b(\phi) = a_bT_{b,\,k}(\phi) + m_{0,\,b}
\end{equation}
in which $a_b$, $m_{0,\,b}$ are free parameters (amplitude and magnitude at peak brightness, respectively), $T_{b,\,k}$ is the $k$th normalized light curve in band $b = \left\{u^*,\,g',\,i',\,z'\right\}$, and phase $\phi$ is determined by:
\begin{equation}
    \phi(t\;|\;P,\;\phi_0) = \frac{t\;\text{modulo}\;P}{P} + \phi_0
\end{equation}
Note the $\phi_0$ in our work was calculated by using the MJD system for the time parameter $t$, and the value of $\phi$ was restricted to $0\leq\phi\leq1$. To get the best-fit values of the free parameters $\left\{a_b,\; m_{0,\,b},\;P,\;\phi_0\right\}$, we minimize the following $\chi^2$ value:
\begin{equation}
    \chi^2 = \sum_{b=u^*,\,g',\,i',\,z'}\sum_{n=1}^{N_{\text{b, obs}}}\left(\frac{m_{b,\,n} - m_b[\phi(t_{b,\,n}|P,\;\phi_0)]}{\sigma_{b,\,n}}\right)^2
\end{equation}
in which $t_{b,\,n}$, $m_{b,\,n}$ and $\sigma_{b,n}$ are the time, apparent magnitude, and photometric uncertainty of the $n$th observation in band $b = \left\{u^*,\,g',\,i',\,z'\right\}$. In the end, a periodogram score, $s_{\text{fitting}} = 1- \frac{\chi^2}{\chi_{0}^2}$, is estimated for the best-fit period to quantify the goodness of fitting. This fitting process is very similar to that which is described in \cite{2017AJ....153..204S}, but the difference is that we applied the normalized light curve model and did not fix the ratio of the light curve amplitudes and colors between the $u^*$ band and the other three bands. The periodogram is calculated with a brute-force search in period space from 0.2 to 0.9~d, with a step size of 0.5~s. For each given period, the fitting of pulsational parameters was done with the Python routine {\tt scipy.optimization.minimize}. Here is our rationale for our adopted template-fitting scheme:
\begin{enumerate}
     \item We decided not to apply the color and amplitude restrictions from the SDSS RR Lyrae templates to our fitting, because the transformations from the SDSS to the CFHT MegaCam photometric system is not well-calibrated for horizontal branch stars like RR Lyrae, especially at the blue end.
    \item The grid interval used in our initial fitting is estimated as 
        \begin{equation}
        \Delta t \sim f\frac{P^2}{T_{\rm obs}} \label{dt}
        \end{equation}
    where $T_{\rm obs}$ is the whole timespan of the NGVS observation, $P$ is an estimated typical period of RR Lyrae and $f$ is an arbitrarily chosen fudge factor to define the desired phase accuracy. The above relationship is derived based on a series of heuristic assumptions. We assume the maximum number of pulsation cycles of RR Lyrae to be $\sim \frac{T_{\rm obs}}{P}$, and the single-cycle pulsation phase shift resulting from the period search grid size to be $\frac{\Delta t}{P}$. The cumulative phase error would therefore be $\frac{T_{\rm obs}\Delta t}{P^2}$, and with a phase accuracy level $f$ selected (we used $f = 0.05$ in this work), we can derive the above equation to estimate the necessary grid interval for the period search.
\end{enumerate}

We understand that the periodogram score defined above is more about the goodness of the fitting and therefore an imperfect indicator of whether an object is an RR Lyrae. We also understand that the chi-square minimization algorithm we have used does not guarantee that the global minimum has been found, mainly because of the extra degrees of freedom we allowed for the $u^*$ band data. The template fitting scheme we have used, however, is a practical choice given our available computational resources, the number of the NGVS photometry epochs, and our multi-band light curve models. Our initial fitting is at least good enough to track the peaks where the global optimal solution resides, and based on the definition of $\Delta t$ shown above, it is reasonable to expect that the periodogram should be mostly smooth within the $\Delta t$ scale.

\subsection{Exclusion of Known Quasars/Active Galactic Nuclei in the NGVS Footprint} \label{subsec:qsos}
After the initial light curve fitting process was completed,  our sample of variable candidates was cross-matched against the known QSOs and AGNs classified with SDSS and XMM-Newton data \citep{2021MNRAS.503.5263Z}. We found 131~matches with spectroscopically confirmed galaxies in SDSS, photometric redshift based galaxies in SDSS, and X-ray luminous QSOs and active galactic nuclei (AGNs) in XMM-Newton. Presumably, the variability we are detecting is related to nuclear activity in these background galaxies. It is reassuring that, for these objects, our light curve fitting algorithm returns low fitting scores: $s_{\rm fitting}<0.3$.

\subsection{Visual Vetting of Light Curves} \label{subsec:visual_lc}
We visually vetted the light curves of the remaining 1554 objects, and classified them into four groups; (a)~366~probable RR~Lyrae candidates, with a relatively high fitting score in the initial round ($s_{\text{fitting}} > 0.5$), good phase coverage, and obvious short-term variability in its unfolded light curve; (b)~71~marginal variable sources; (c)~615~suspected QSO/AGN contaminants with a low fitting score ($s_{\text{fitting}} < 0.3$), that display clear long-term variability through multiple observing seasons with similar photometric trends across the different bands; and (d)~502~sources that appear to be non-variable that are likely to be affected by systematic errors (e.g., saturation, cosmic rays, charge bleeds from neighboring bright stars, effect of variable seeing on close neighbors, detector artifacts, and other possible calibration errors in our time-series photometry). Representative examples of these four categories are shown in Figure~\ref{fig:Classes}. All objects classified as type (a) and (b) are further analyzed with a more detailed light curve fitting process.

\subsection{Zoom-in Template Fitting}\label{subsec:zoom}
We perform a zoom-in fitting to all variable candidates around the 10~periodogram peaks shown in their initial fitting results, in order to avoid missing the global minimum of the periodogram as the ideal $\Delta t$ interval in 
Equation~\ref{dt} may vary for RR~Lyrae with different periods and observation baselines. For each peak $P_{\text{peak}}$ in the initial periodogram, we search its nearby $\pm 0.5$~s range with a zoom-in fitting grid interval $\Delta t_{\text{zoom}} = 0.02$~s. The period value that corresponds to the highest fitting score in the zoom-in fitting process is then recorded as the best-fit period $P_{\text{best}}$. In most cases, the $P_{\text{best}}$ value occurs around the tallest peak in the initial fitting, as shown in the upper panels of Figure~\ref{fig:periodo}. However, we also found cases where $P_{\text{best}}$ occurs around suboptimal peaks in the initial fitting, as shown in the lower panels of Figure~\ref{fig:periodo}, suggesting that we were slightly overestimating the ideal $\Delta t$ interval in the initial fitting. The best-fit period $P_{\text{best}}$ and its corresponding fitting parameters (best-fit templates, pulsational parameters, and fitting score) were then used for further vetting process. Each variable candidate is assigned the type (RRab or RRc) of the best-fit SDSS Stripe~82 RR~Lrae template.

\subsection{Robustness Tests Using Known RR Lyrae in the NGVS Footprint} \label{subsec: PS1Test}

\begin{figure}
    \centering
    \includegraphics[width=0.45\textwidth]{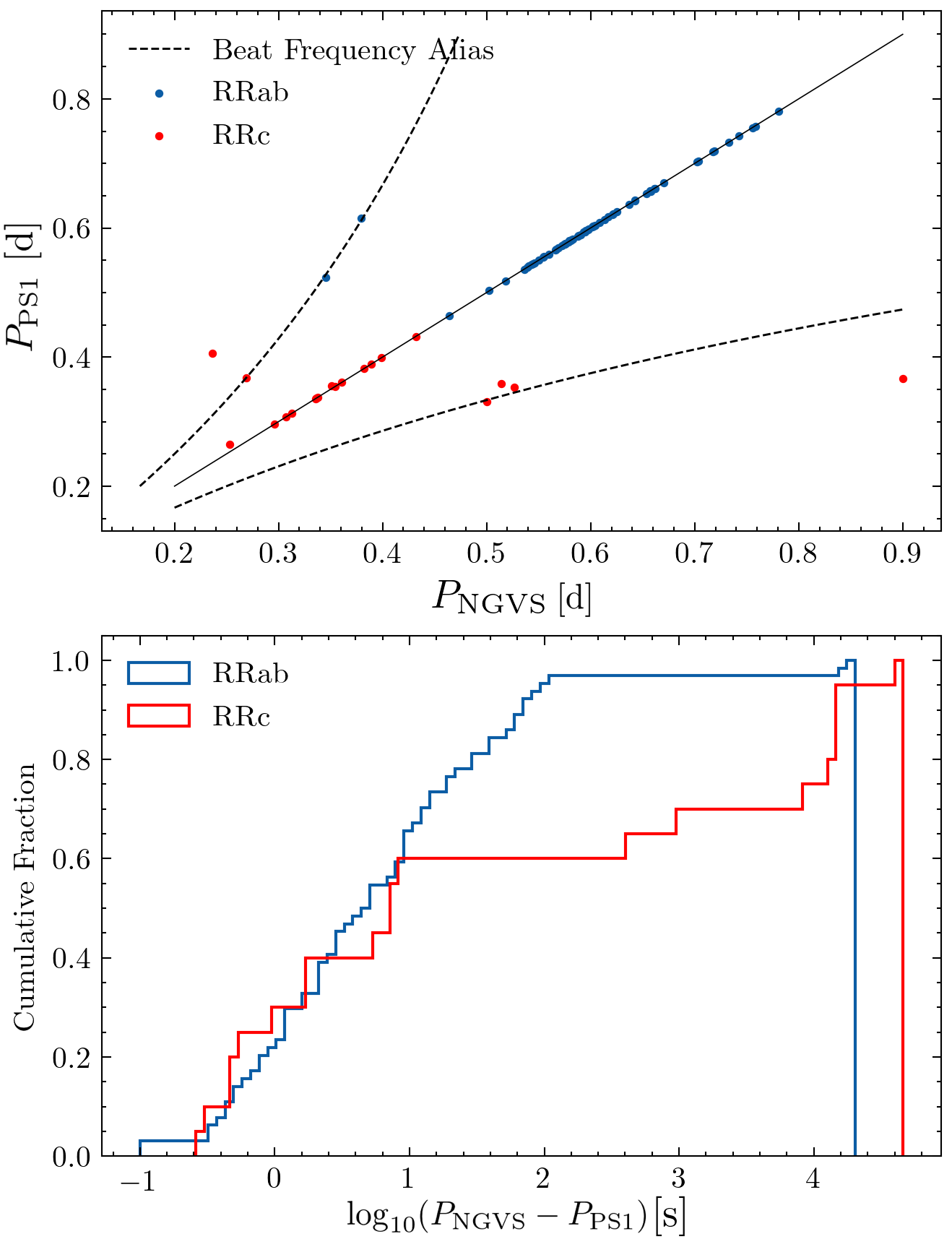}
    \caption{Accuracy, precision, and robustness of the RR Lyrae period estimations obtained using our multi-band light-curve template fitting of NGVS light curves. The top panel compares periods estimated by \citep{2017AJ....153..204S} with those measured from NGVS data using our multi-band template fitting. The dashed lines show the one-day ($N=\pm1$) beat frequency aliases.$^1$ The bottom panel quantifies the precision of the period recovery: the period is accurately recovered (i.e., within $\pm1$~min) for 82\% of RRab and 60\% of RRc stars.}
    \label{fig:ppvs}
\end{figure}

\begin{figure*}
\gridline{\fig{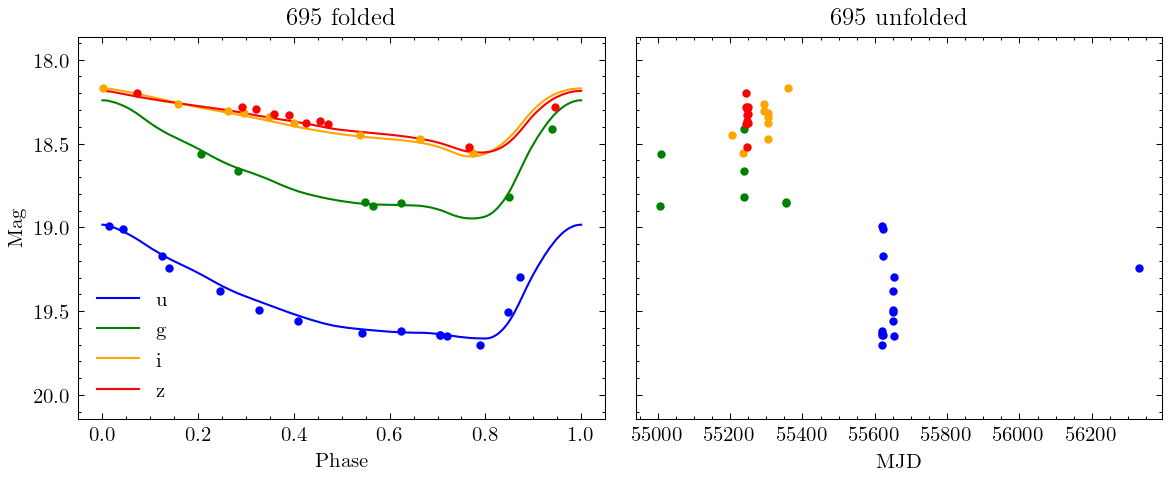}{0.45\textwidth}{(a)}
          \fig{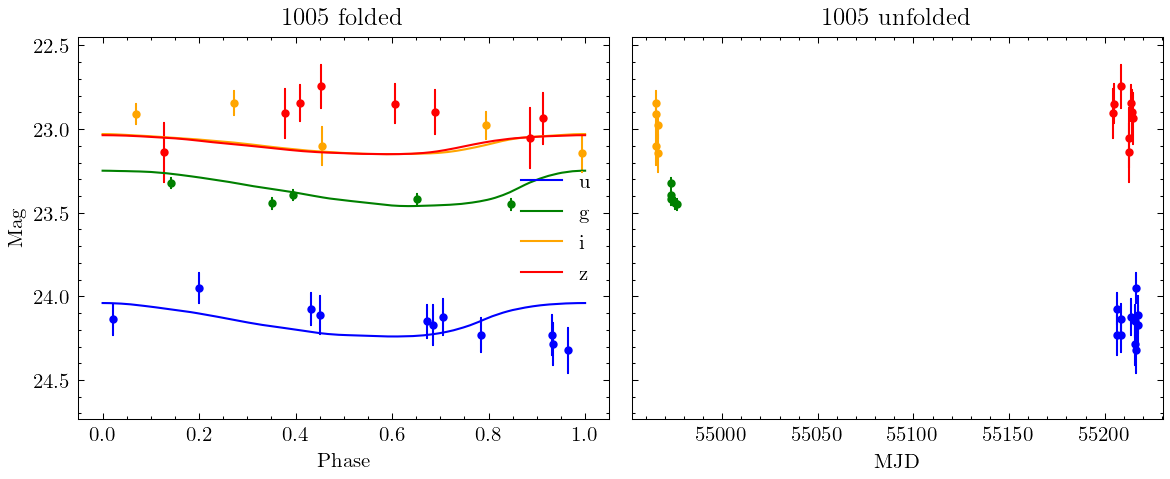}{0.45\textwidth}{(b)}
          }
\gridline{\fig{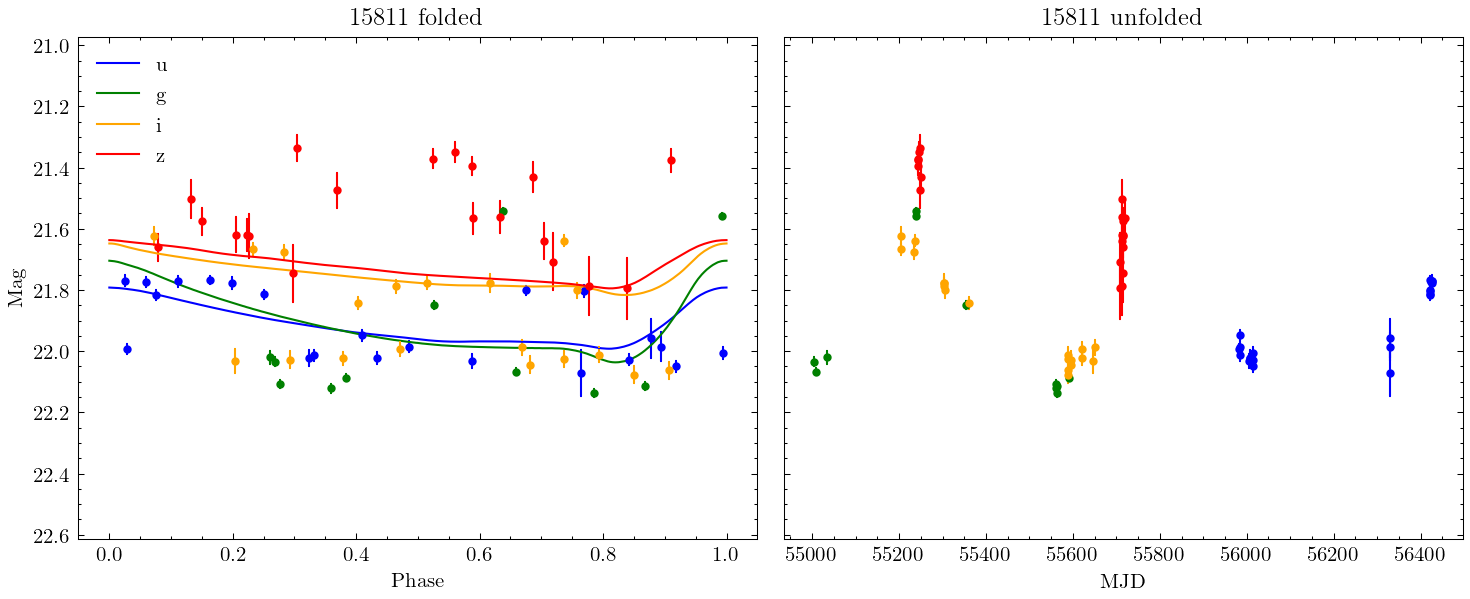}{0.45\textwidth}{(c)}
          \fig{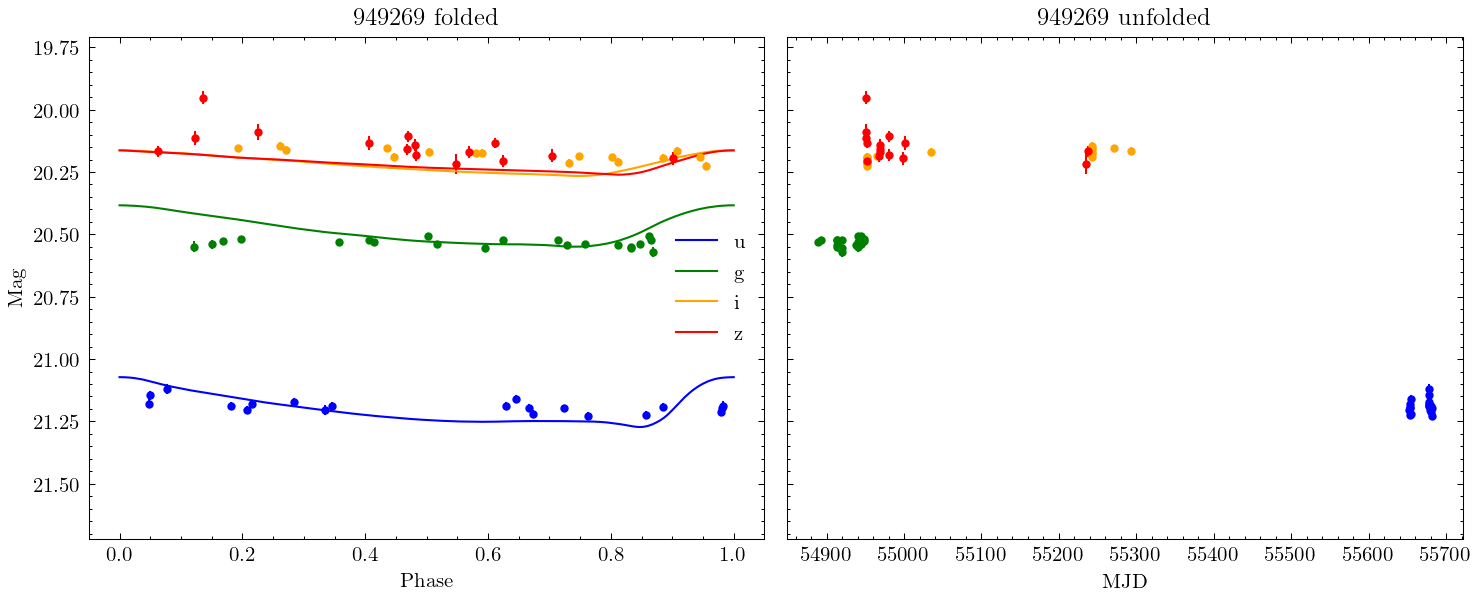}{0.45\textwidth}{(d)}
          }
\caption{An example of each of the four types of variable candidates after the first round of fitting, with their folded and unfolded light curves:
(a) definite RR Lyrae: well-fit folded light curve and clear short-term variability in its unfolded light curve; (b) marginal variables: low score fitting with clear short-term variability; (c) definite QSOs, low score fitting with clear long-term variability; and (d) non-variables. All variables are visually classified into these four categories, and objects marked as category (a) and (b) have been further analysed via expanded fitting (see \S\,\ref{subsec:zoom}).
\label{fig:Classes}}
\end{figure*}

The result of applying this fitting procedure to the unfolded light curves of the 84 PS1 RR Lyrae candidates is illustrated in Figure \ref{fig:ppvs}. Our multi-band template fitting method accurately measures periods for $91\%$ of RR Lyrae stars ($97\%$ of RRab and $70\%$ of RRc stars). The period is recovered to within 1 sec for $73\%$ of RR Lyrae stars. For all fittings with score $s_{\text{fitting}} > 0.93$, the period differences are within $3$ min. Note that if the period fitting returns a discrepant value, this can predominately be attributed to one-day beat frequency aliasing.\footnote{The beat frequency is: $f_{\text{beat}} = f_{\text{true}} + Nf_{\text{sample}}\;( N = \pm1, \pm2, ...)$. For ground based surveys like NGVS, $f_{\text{sample}} \approx1$~d$^{-1}$. The one-day beat period (in d) is therefore given by: $P_{\text{beat}} = \frac{P_{\text{true}}}{1+NP_{\text{true}}}$, as shown by the curved dashed lines in the upper panel of Figure \ref{fig:ppvs} (for $N = \pm1$).}

The differences in the fitted pulsational parameters are remarkably small considering that we are using completely different observational data, as well as sparsely sampled observations compared with the PS1 study.

We also note that we recover the halo RR Lyrae star discovered by \citet{1989AJ.....98.1648C} near M49 (NGC~4472), which was the most distant MW halo star observed at that time. The period that we fit is within 15~s (0.03\%) of that measured by the original discoverers. 

\subsection{Visual Vetting of Images} \label{subsec:visual_img}
We visually inspected the stacked and single exposure $g'$ band images of all variable candidates with $s_{\text{fitting}} > 0.9$ to guard against a couple of factors could lead to false positives in our RR~Lyrae search. First, the NGVS footprint contains a large number of star-forming Virgo cluster galaxies whose photometrically variable supergiants could masquerade as MW halo RR~Lyrae \citep[e.g., the study of NGC~4535 by][]{2018A&A...618A.185S}. Second, the combination of variable seeing and sky subtraction errors for point sources that are located in complicated backgrounds (e.g., within the dust lanes/spiral arms of star-forming galaxies in the Virgo cluster) can compromise the fidelity of the stacked-image and time-series photometry. We found 8~objects with complicated backgrounds and noticed that their scores are in the range: $0.9 < s_{\text{fitting}} < 0.93$. This led us to use an RR~Lyrae selection criterion of $s_{\text{fitting}} > 0.93$.

We searched for our faintest NGVS RR~Lyrae candidates ($g' > 21.5$) in the SMOKA archive to check if they had been observed with the Subaru 8-m telescope, and found 11 matched stars. The single exposure Hyper Suprime-Cam HSC-$g$ band images of these objects, obtained between 2014 and 2018, show that none of them are transients. A similar determination was made for 2~additional NGVS RR~Lyrae candidates that were matched against {\it Hubble Space Telescope\/} Advanced Camera for Surveys images in the MAST archive.
\begin{figure*}
    \centering
    \includegraphics[width=1.0\textwidth]{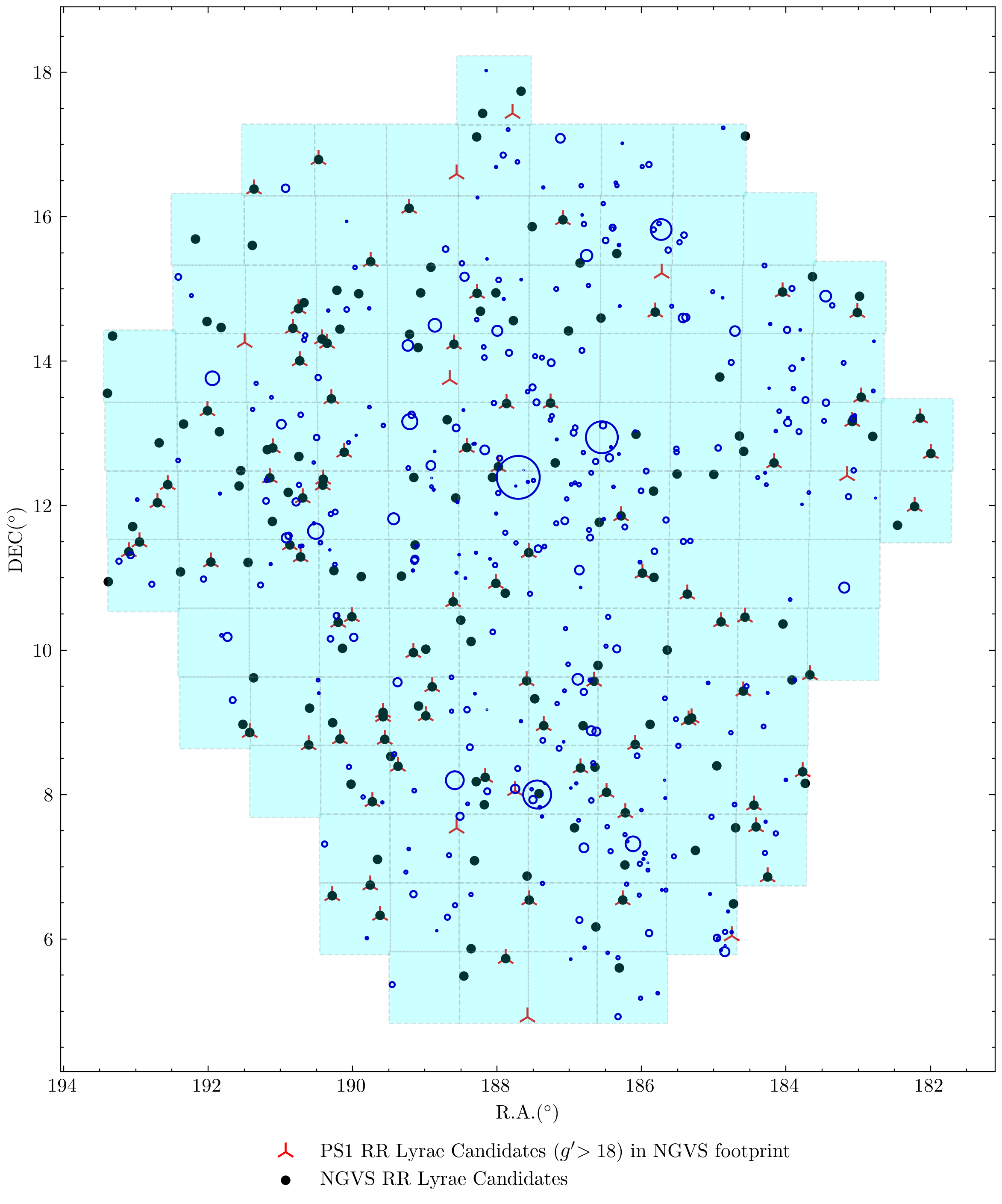}
    \caption{Sky distribution of all 117~NGVS pointings, each $1^\circ\times1^\circ$ and with slight overlap between adjacent pointings (cyan squares), PS1 RR Lyrae candidates with $g' > 18$~mag (red three-pronged symbols), and NGVS RR Lyrae candidates (filled black dots). The 350~brightest galaxies within the NGVS footprint are delineated with blue circles, the radii of which are equal to $5\;r_{\text{eff}}$. The $g' > 18$~mag limit is applied to the PS1 sample because photometric saturation occurs at this brightness level for NGVS. There are 94 PS1 RR Lyrae candidates with $g' > 18$~mag in the NGVS field, of which 84 are classified as RR Lyrae in this work (see \S\,\ref{sec:clr_clr_sel}).}
    \label{fig:PS1_caompare_spatial}
\end{figure*}

\begin{figure*}
\gridline{\fig{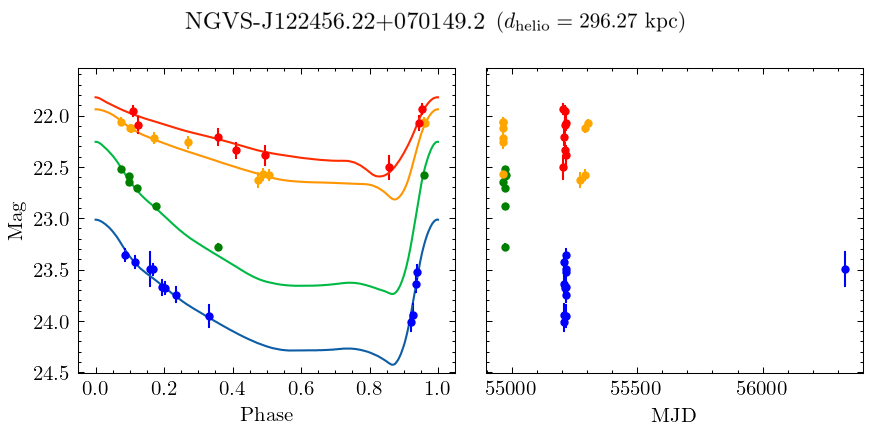}{0.5\textwidth}{(a)}
        \fig{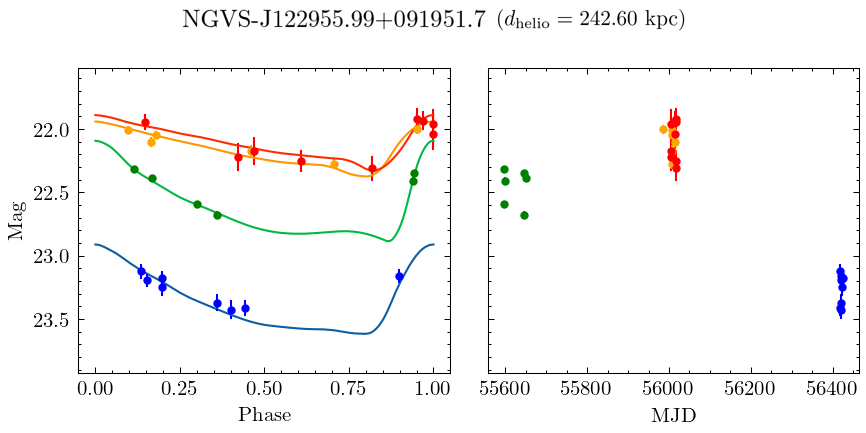}{0.5\textwidth}{(b)}
          }
\gridline{\fig{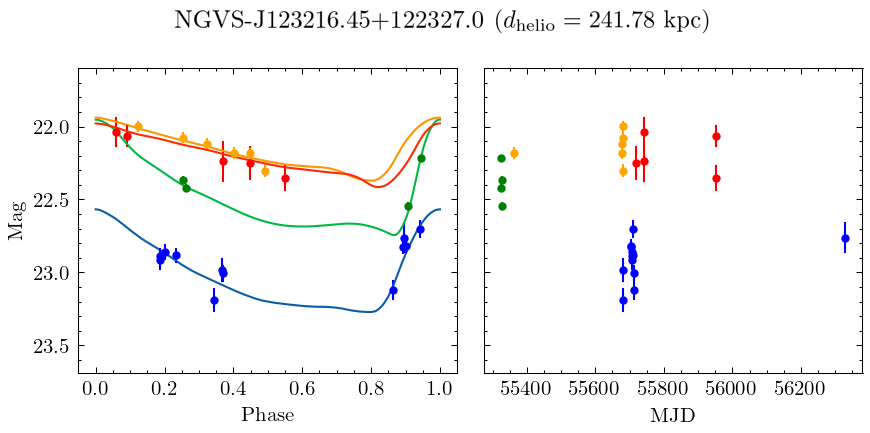}{0.5\textwidth}{(c)}
          \fig{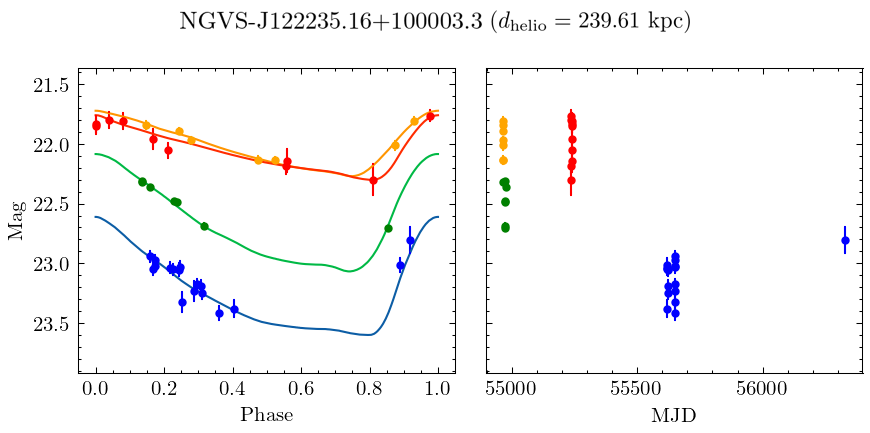}{0.5\textwidth}{(d)}
          }
\gridline{\fig{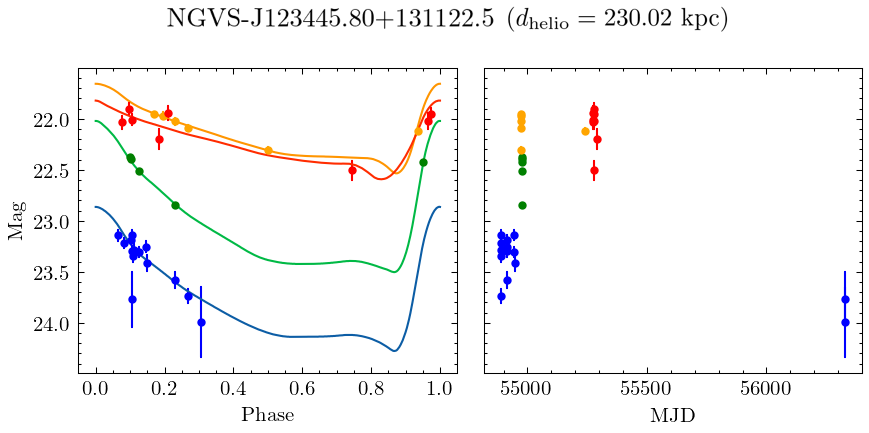}{0.5\textwidth}{(e)}
          \fig{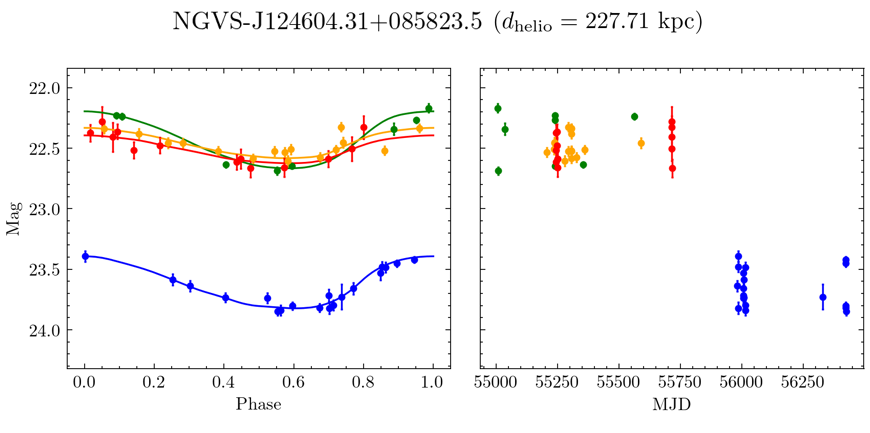}{0.5\textwidth}{f}
          }
\caption{Light curves of the six most distant NGVS RR Lyrae candidates, each labeled with their object ID and heliocentric distance. The color scheme is same as in Figure~\ref{fig:Classes}.}
    \label{fig:LC_exp}
\end{figure*}

\begin{figure*}
\gridline{\fig{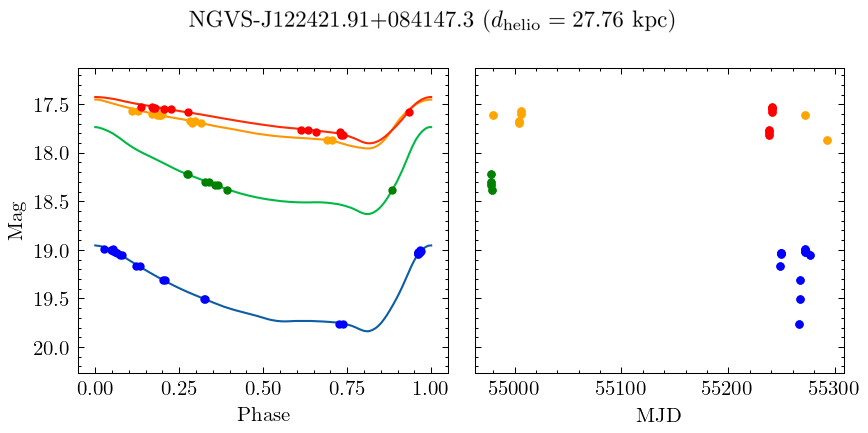}{0.5\textwidth}{(a)}
        \fig{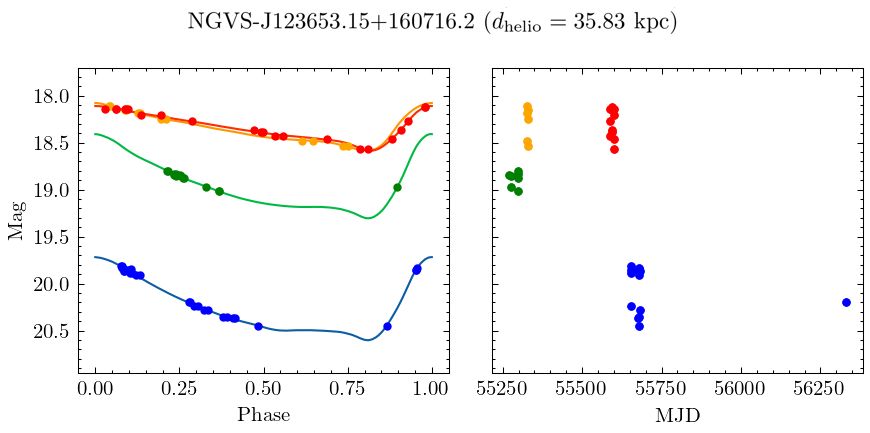}{0.5\textwidth}{(b)}
          }
\gridline{\fig{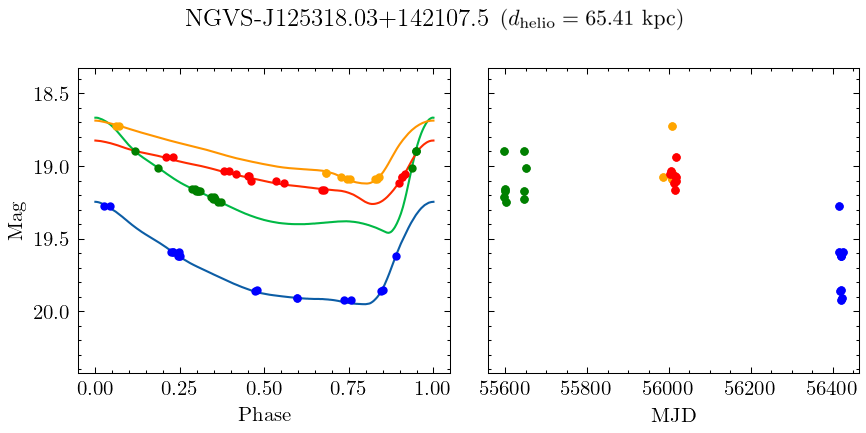}{0.5\textwidth}{(c)}
        \fig{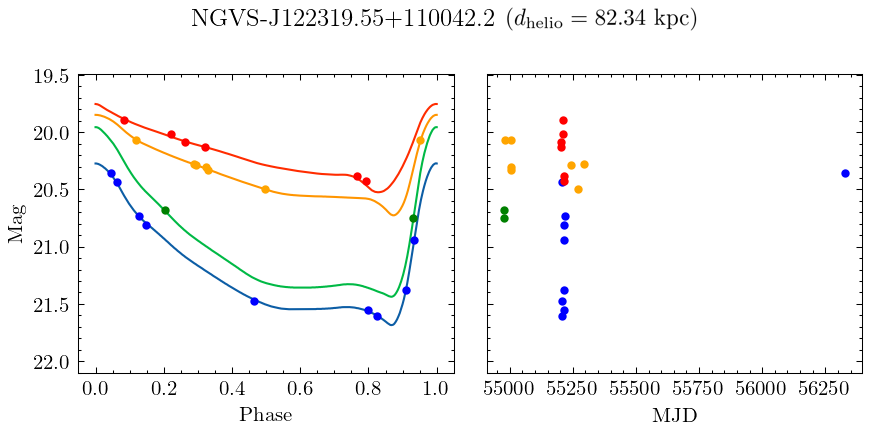}{0.5\textwidth}{(d)}
          }
\gridline{\fig{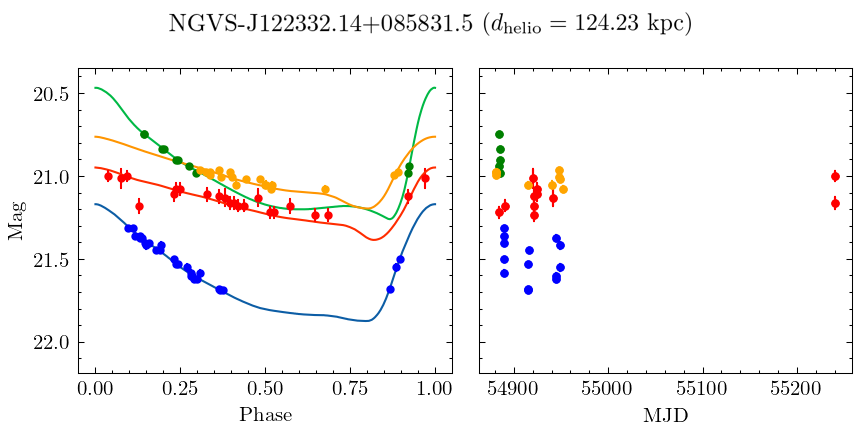}{0.5\textwidth}{(e)}
        \fig{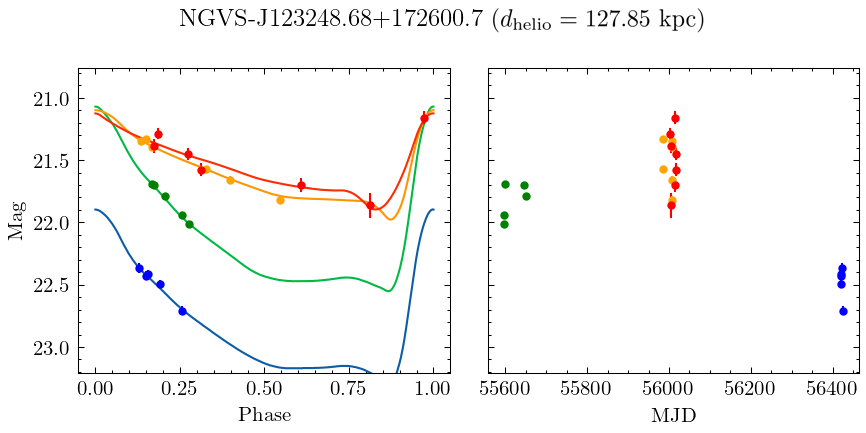}{0.5\textwidth}{(f)}
          }
\caption{Light curves of six representative NGVS RR Lyrae candidates at different heliocentric distances within 150~kpc, each labeled with its object ID and heliocentric distance. The color scheme is the same as in Figure~\ref{fig:Classes}.}
    \label{fig:LC_exp_2}
\end{figure*}

\subsection{The RR Lyrae Sample}
In total, we have found 180 RR Lyrae candidates in the NGVS field, of which there are 139 RRab and 41 RRc. The light curve fittings of all RR~Lyrae candidates passed the threshold periodogram score value of $s_{\text{fitting}}> 0.93$ and were checked visually. The best-fit pulsation parameters and other information are available in a machine-readable table, with the column definition and the first few table entries shown in Table~\ref{tab:RR catalog}. For each NGVS RR Lyrae candidate, we integrate over its best-fit light curve---based on its best-fit pulsational parameters: period, amplitude, phase, and light curve shape---to calculate its flux-averaged apparent magnitude $\langle{m_b}\rangle$ in each band (where $b=u^*$, $g'$, $i'$, and $z'$). The extinction was corrected using the dust map in \cite{2023ApJ...958..118C} and the $u^*g'i'z'$ extinction coefficients calibrated by \cite{2014ApJS..210....4M}.

All of our RR Lyrae candidates are projected more than 5\,$r_{\text{eff}}$ from the center of the closest (in projection) bright Virgo cluster galaxy, with two exceptions: (1) one is projected within 1\,$r_{\text{eff}}$ of the center of the bright and smooth giant elliptical galaxy M49 in the Virgo cluster, and is an RR~Lyrae that was independently discovered by \citet{1989AJ.....98.1648C}; and (2) the other is projected 4.8\,$r_{\text{eff}}$ from the center of the early-type spiral galaxy NGC~4492 in the Virgo cluster, and is an RRab that was independently discovered by PS1 \citep{2017AJ....153..204S}.

Although all of our RR~Lyrae candidates have been visually inspected, it would nevertheless be useful to obtain additional photometry to confirm the RR~Lyrae classification of those candidates that are poorly sampled in one or more bands. In Figure~\ref{fig:PS1_caompare_spatial}, we show a sky map of our NGVS RR~Lyrae candidates.

\begin{deluxetable*}{ccccccccccccccc}[!t]
\tablenum{2}
\tablecaption{Catalog of NGVS RR Lyrae Stars\label{tab:RR catalog}}
\tablewidth{0pt}
\tablehead{
\colhead{NGVS ID} & \colhead{R.A.} & \colhead{Decl.} & \colhead{$\langle u^* \rangle$} & \colhead{$\left \langle g' \right \rangle$} & \colhead{$\left \langle i' \right \rangle$} &
\colhead{$\left \langle z' \right \rangle$} &
\colhead{$d_{\rm helio}$} & \colhead{$u^*_{\rm amp}$} & \colhead{$g'_{\rm amp}$} & \colhead{$i'_{\rm amp}$} & \colhead{$z'_{\rm amp}$} & \colhead{Period} & \colhead{Type} & \colhead{$\phi_0$}\\
\colhead{ } & \colhead{[deg]} & \colhead{[deg]} & \colhead{[mag]} & \colhead{[mag]} & \colhead{[mag]} & \colhead{[mag]} & \colhead{[kpc]} & \colhead{[mag]} &\colhead{[mag]} &\colhead{[mag]} & \colhead{[mag]} &\colhead{[d]} &\colhead{}
}

\startdata
NGVSJ121701.35$+$065134.9 & 184.255618 &  6.8597015 & 19.30 & 18.56 & 18.29 & 18.22 & 36.14 & 1.21 & 0.97 & 0.51 & 0.46 & 0.7194 & ab & 0.36
\enddata
\tablecomments{The above table shows the column information of our NGVS RR Lyrae catalog. The full catalog, together with the multicolor light curves, are available online at: https://www.canfar.net/citation/landing?doi=24.0002.}
\end{deluxetable*}


\subsection{Distance Determination}\label{sec:dist_det}
We directly adopted the $i'$-band $P$-$L$ relation in \cite{2017AJ....153..204S}, which was calibrated with PS1 RR Lyrae data, since there are no known MW star clusters or satellite dwarf galaxies in the NGVS footprint for an independent calibration. We calculated the distances of our RR Lyrae candidates with their best-fit $i'$-band flux-averaged magnitude and the $P$-$L$ relation in \cite{2017AJ....153..204S}, where the absolute $i'$-band magnitude is estimated as: 
\begin{equation}
    M_{i'} = -1.77\log_{10}{(P/0.6)} + 0.46
\end{equation}
and for RRc stars their periods are fundamentalized before calculating their absolute magnitudes \citep{2009Ap&SS.320..261C}:
\begin{equation}
    \log_{10}{P_{F}} = \log_{10}{P} + 0.128
\end{equation}
The uncertainty of the distance estimation calculated by the $i'$-band $P$-$L$ relation in \cite{2017AJ....153..204S} was reported to be $\sigma_{M_{i'}} = 0.06~(\text{random}) \pm 0.03~(\text{systematic})$. Here, we validate that the uncertainty of the distance estimation in our study can be constrained from the PS1 result. First, the differences between the flux-averaged $i$-band magnitudes in our study and in PS1 for overlapping RR Lyrae stars are at a remarkably low level ($< 0.01$ mag for non-aliasing cases). Second, the error induced by the ``fundamentalization'' of RRc star periods (by Catelan 2009) is smaller than 0.002 mag. Third, the $P$-$L$ relation has only a weak dependence on metallicity ($Z$). Sesar et al.\ claimed that their period-luminosity-metallicity ($P$-$L$-$Z$) relation has a scatter of $\sigma_{\rm DM} = 0.09$ with a reference metallicity of $\rm[Fe/H] = -1.5$. While faint stars in the Virgo direction reside in the outer halo and could be metal-poor, the dependence of absolute magnitude on metallicity is weak. Even for the extreme metal-poor case, the resulting error in the distance modulus is $\sigma \sim 0.1$ mag, corresponding to a $5\%$ error in distance even at 300 kpc.

\subsection{Completeness Tests}\label{subsec:completeness}
We estimate the completeness of our NGVS RR~Lyrae sample by using a large synthetic RR~Lyrae data set that mimics the cadence and photometric errors of the NGVS. In order to generate the light curve of a synthetic RR~Lyrae with a flux-averaged magnitude $\langle g'_{\text{synth}}\rangle$, the set of $ugiz$ light curve templates associated with a random `parent star' with flux-averaged magnitude $\langle g_{\text{parent}}\rangle$ is selected from the \citet{2010ApJ...708..717S} catalog of 483 SDSS RR~Lyrae. The pulsational parameters of the parent star are adopted: period and $ugiz$ amplitudes. The difference $\Delta\langle g'_{\text{synth}}\rangle = \langle g'_{\text{synth}}\rangle - \langle g_{\text{parent}}\rangle$, which is equal to the difference in distance moduli between the synthetic and parent RR~Lyrae, is then added to the light curves of the parent star in all four bands, and a random initial phase is assigned. The mock NGVS observation of this synthetic RR~Lyrae is simulated by randomly selecting a point source in our NGVS time-series catalog, importing its epochs in the four bands, and calculating the folded phases and magnitudes of each simulated NGVS observation based on the light curve of the synthetic star. Next, this time-series data set is converted from the SDSS $ugiz$ system to the NGVS $u^*g'i'z'$ system by applying the inverse of the photometric calibration procedure mentioned in \S\,\ref{subsec:stacked_phot} \citep[for details, see][]{2008PASP..120..212G}. Finally, we incorporate realistic photometric errors in order to create a synthetic RR~Lyrae time-series data set.

In total, we generated 7200 synthetic RR~Lyrae (4800 RRab and 2400 RRc) time-series data sets that sample the full range of periods, amplitudes, and light curve shapes of the 483 SDSS Stripe 82 RR~Lyrae, but shifted to the apparent magnitude range $20.5\leq \langle g' \rangle \leq 24.5$. This mock RR~Lyrae sample was analysed using the same light curve template fitting process described in \S\S\,\ref{subsec:initial}--\ref{subsec:zoom}. A synthetic RR~Lyrae is considered recovered if it satisfies the multi band variability criteria and fitting score threshold, and the best-fit period is within $\pm5$~min of the period of the parent star.

\begin{figure}
    \centering
    \includegraphics[width=0.7\textwidth]{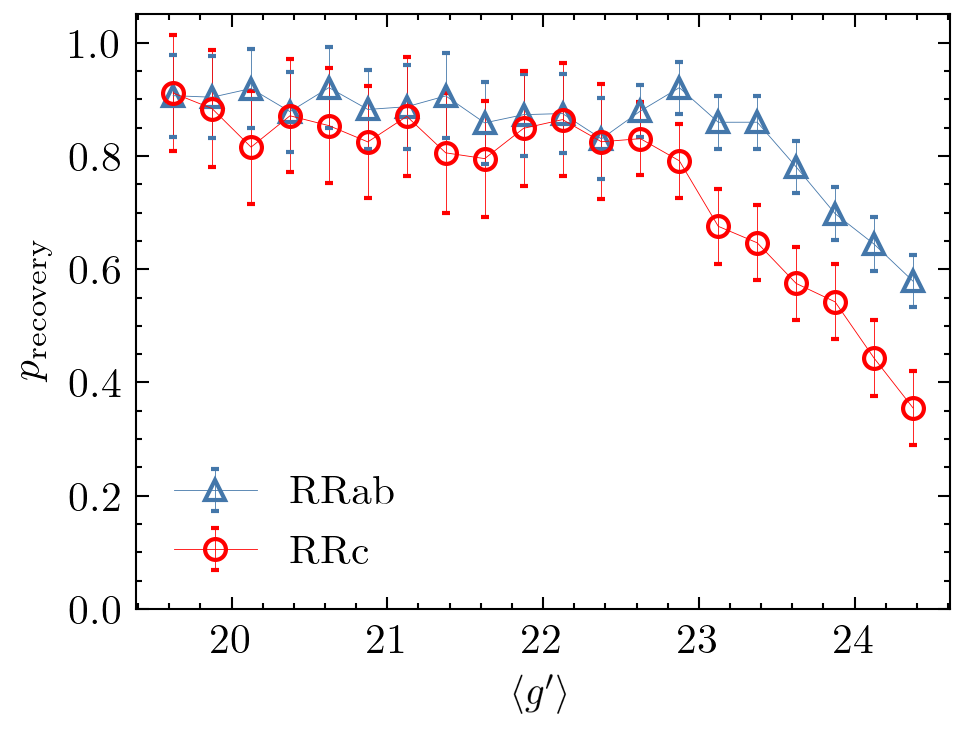}
    \caption{Recovery fraction of RRab and RRc stars (blue triangles and red circles, respectively) in NGVS as a function of flux-averaged $\langle g'\rangle$ magnitude, based on integration over the full (known) light curve. The error bars shown represent Poisson fluctuations. These completeness statistics are based on tests using synthetic RR~Lyrae stars, as described in \S\,\ref{subsec:completeness}.}
    \label{fig:completeness}
\end{figure}

The resulting completeness results are shown in Figure~\ref{fig:completeness}. For RRab stars, the completeness is in the range 85\%--90\% at bright magnitudes, and drops from 88\% to 58\% over the range $\langle g'\rangle = 23.4$--24.45. For RRc stars, the completeness is in the range 80\%--85\% at bright magnitudes, and drops from 80\% to 36\% over the range $\langle g'\rangle = 22.8$--24.4. The smaller pulsation amplitudes, shorter periods, and more symmetric light curve shapes of RRc relative to RRab makes it harder for the light curve fitting procedure to recover them. The apparent magnitude corresponding to the 85\% recovery rate for RRab is 1.8~mag deeper for NGVS than HiTS \citep{2018ApJ...855...43M} and 1.3~mag deeper for NGVS than DES \citep{2021ApJ...911..109S}. The NGVS RRab completeness curve shown in Figure~\ref{fig:completeness} is used to correct the raw radial density profile of MW halo RR~Lyrae, as described in \S\,\ref{subsec:density}. The RRab completeness level is assumed to be constant at 90\% (average of the first three bins in Figure~\ref{fig:completeness}) from $\langle g' \rangle=19.0$--20.5, a range over which there is no saturation and the photometric accuracy is high. It is worth mentioning that these completeness estimates do not take into account the areal completeness of the NGVS survey. However, given that the 5\,$r_{\text{eff}}$ regions of brightest 350~galaxies combined cover only 1.1\% of the NGVS survey footprint (shown in Figure~\ref{fig:PS1_caompare_spatial}), the effect of the Virgo cluster background on our completeness estimates is negligible.

In Figure \ref{fig:LC_exp}, we demonstrate the robustness of our NGVS RR~Lyrae sample by showing the folded and unfolded light curves for the 6~most distant (faintest) candidates. The quality of the template fits for these
6~distant RR~Lyrae is comparable to that of brighter NGVS RR~Lyrae (Figure~\ref{fig:LC_exp_2}) and better than that of comparably distant/faint HiTS and DES RR~Lyrae candidates \citep[][e.g., see their Figure~3]{2018ApJ...855...43M,2021ApJ...911..109S}. The properties of the full sample of 180 NGVS RR Lyrae are presented in Table~\ref{tab:RR catalog}. 

\section{Results and Discussion} \label{sec:results}


\subsection{Reliable RR Lyrae Detections in Outer Halo}
\begin{figure*}[!htb]
    \centering
    \includegraphics[width=0.9\textwidth]{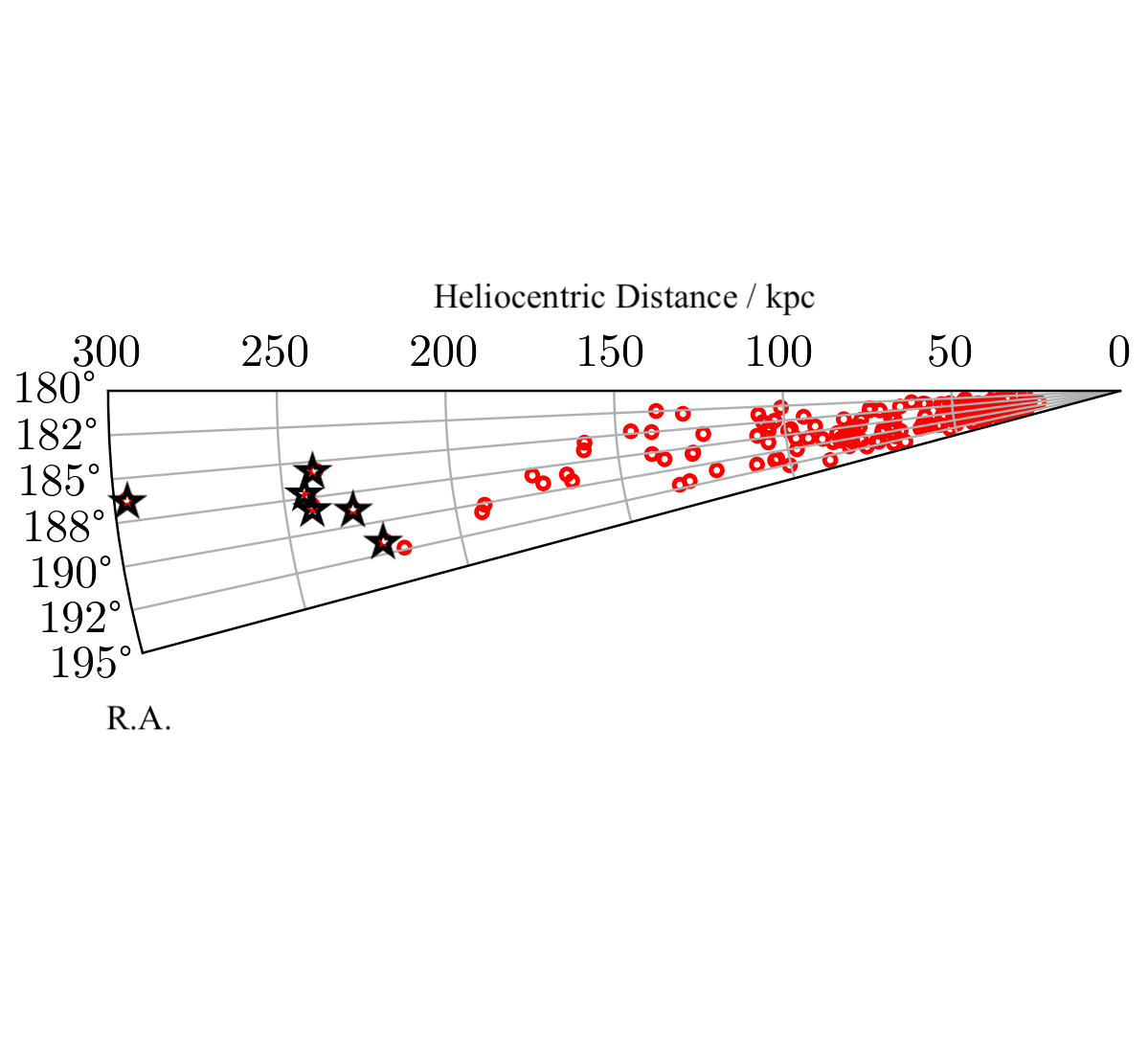}
    \caption{Radial plot of heliocentric distances as a function of right ascension (R.A.). The folded and unfolded light curves of the six most distant RR Lyrae are shown in Figure \ref{fig:LC_exp}.
    }
    \label{fig:wedge}
\end{figure*}

Our NGVS sample of 180 RR~Lyrae candidates roughly doubles the number of known RR~Lyrae in this region of the sky: PS1 reported 94 RR~Lyrae, whereas the additional RR~Lyrae we have found in the NGVS database are, for the most part, fainter than the detection limit of PS1 (\S\,2.2). The primary advantage of our NGVS data set over HiTS \citep{2018ApJ...855...43M} and DES \citep{2021ApJ...911..109S} is significantly greater single-epoch photometric precision/depth (Figure~\ref{fig:chi_std}). As a result, we expect that
our sample of NGVS RR~Lyrae candidates represents a more robust detection and reliable characterization of the most distant known RR~Lyrae in the MW halo. Our NGVS sample of RR~Lyrae candidates includes 39 with $d_{\rm helio}>100$~kpc and 7 with $d_{\rm helio}>200$~kpc. We show the full sample of RR Lyrae stars as a function of distance and RA in Figure~\ref{fig:wedge}; the six most distant objects whose light curves are shown in Figure~\ref{fig:LC_exp} are marked by the bold black star symbols.

\subsection{Radial Density Profile}
\label{subsec:density}
\begin{figure*}
    \centering
    \includegraphics[width=0.85\textwidth]{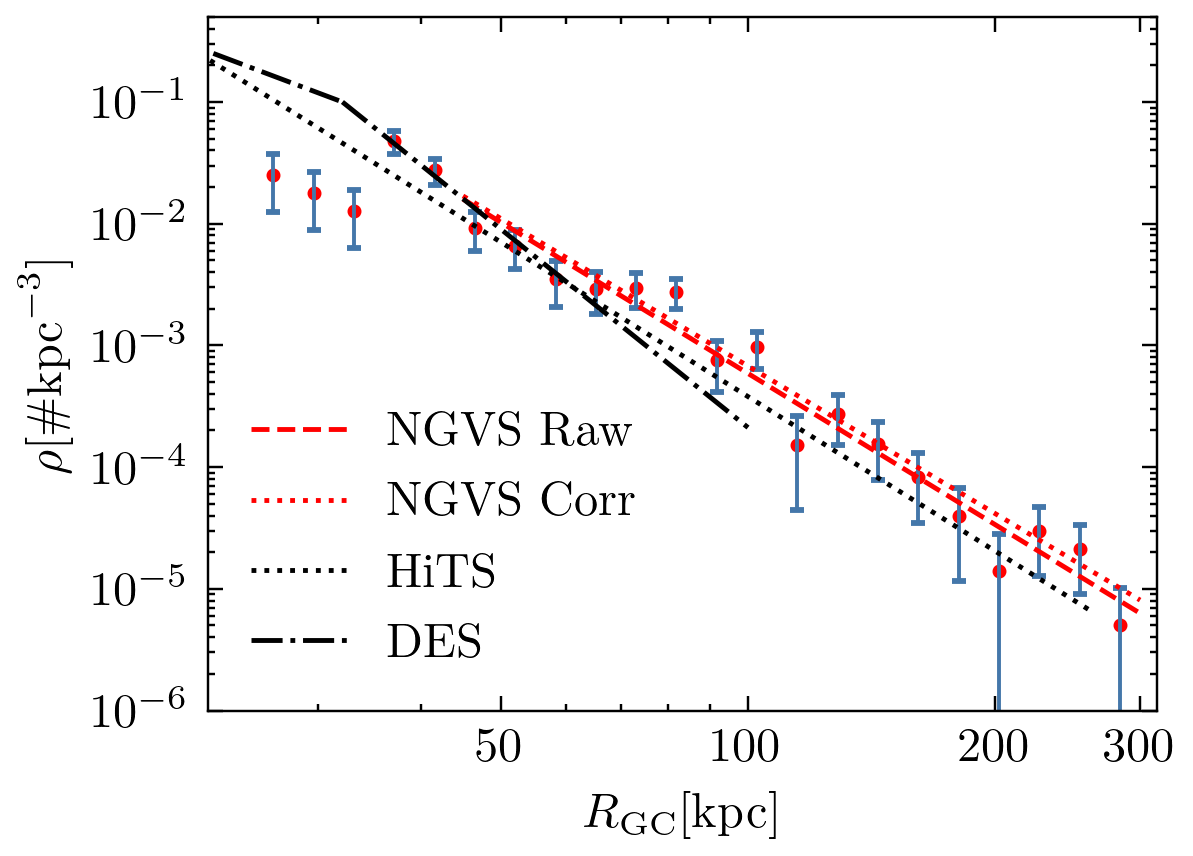}
    \caption{The black dashed line shows the best-fit radial density profile of our NGVS RR Lyrae sample, based on a spherical power-law halo density model. The figure shows no clear cutoff beyond 250~kpc.  Power-law fits to two density profiles {\bf based on our NGVS RR~Lyrae sample}, built with and without the correction for our RRab detection completeness, are shown with a red dotted line (slope $n = -4.09 \pm 0.10$) and a red dashed line (slope $n = -4.16 \pm 0.11$), respectively. We only consider Poisson errors, and the fitting starts from 45~kpc. It is currently the most robust density profile estimation of the outer stellar halo of the MW ($R_{\rm GC}>100$ kpc).}
    \label{fig:density}
\end{figure*}

We construct the MW stellar halo number density radial profile $\rho({R_{\rm GC}})$ in this direction of the sky based on our NGVS RR Lyrae sample, where $R_{\rm GC}$ is the distance to the center of the Galaxy. For this number density calculation, we consider only the 140~RRab candidates identified in our study, spread over an area of 104~deg$^2$ in the direction of the Virgo cluster. The Galactocentric radius $R_{\rm GC}$ is calculated as follows:
\begin{equation}
    R_{\rm GC}^2 = (R_\odot - d_{\rm helio}\cos b \cos l)^2 + (d_{\rm helio}\sin b)^2 + (d_{\rm helio}\cos b \sin l)^2
\end{equation}
where $d_{\rm helio}$ is the heliocentric distance, $l$ and $b$ are the Galactic latitude and longitude of each star, respectively, and $R_{\odot}$ is the distance from the Sun to the Galactic center. We adopt an $R_{\odot}$ value of 7.9~kpc \citep{2020PASJ...72...50V}. We fit the radial number density profile of our RRab sample with a single power-law model: $\rho = \rho_{0}(R_{\rm GC}/R_0)^{n}$ for a spherical halo, in which $\rho_0/R_0^{-n}$ is the normalization constant. The bins are evenly spaced on a log scale as shown in Figure~\ref{fig:density}, and the fitting was weighted with Poisson errors of the density values in each bin. We restrict the power-law fit to $R_{\rm GC} > 45$~kpc because: (1) as these distances, the NGVS sample is unaffected by saturation; and (2) this avoids the well-known Sagittarius stream and Virgo Overdensity substructure along this line of sight \citep[e.g.,][]{2019ApJ...886...76D}. Various observational studies have reported the existence of a break in the halo density radial profile at $R_{\rm GC}\sim20$--35~kpc \citep[e.g.,][]{2014ApJ...781...22Z, 2015ApJ...809..144X} and have therefore adopted a broken power-law model to mark the transition between the inner (in-situ) halo and outer (ex-situ) halo. For this work, however, we only probe the outer halo beyond 45~kpc in the Virgo direction, so use only a single power-law model.

We detect the existence of RR~Lyrae stars out to $R_{\rm GC}\sim300$ kpc without any clear evidence of a break in the power-law density profile out to this radius. This corroborates similar results in the DES Y6 RR~Lyrae catalog \citep{2021ApJ...911..109S}, and the most distant Mira stars recently presented by \citet{2022A&A...660A..35N} with VVV survey data. Also, the Galactic splashback radius, namely the edge of the MW halo, is predicted to be $\sim0.8r_{200\rm m} =  290\pm 61$ kpc by \citet{2020MNRAS.496.3929D}, which corresponds to the Galactocentric radii of the most distant RR Lyrae presented in our work.

The slope of the number density radial profile of the Galactic outer stellar halo has been studied using different stellar tracers, including RR~Lyrae, K giants, blue horizontal branch/blue straggler stars, and A stars \citep[see][]{2018Hernitschek,2018ApJ...855...43M,2021ApJ...911..109S}. The derived density slopes of the outer halo lie in a wide range, between $n = -3.8\pm0.1$ to $n=-5.4\pm0.1$. Our result, $n = -4.09\pm0.10$, is consistent with the $n = -4.17^{+0.18}_{-0.20}$ slope obtained by \cite{2018ApJ...855...43M} using RR~Lyrae stars, and is within the range of other slope estimates. It should be emphasized that our sample of NGVS RR~Lyrae and the DES Y6 RR~Lyrae are currently the only two probes of the outer stellar halo out to $R_{\rm GC}\sim300$~kpc. However, \citet{2021ApJ...911..109S} opted to only fit the DES RR~Lyrae stellar density radial profile in the distance range $30<d_{\rm helio}<100$~kpc, and obtained a steep outer halo slope of $n=-5.42\pm0.13$. The density profile of their RR~Lyrae candidates is significantly shallower than this power-law slope beyond 100~kpc (see their Figure~11). Their choice to limit the power-law fit to $d_{\rm helio}<100$~kpc was motivated by two factors: (1)~they believe that there is unaccounted for QSO contamination in their sample of RR~Lyrae candidates at large distances; and (2)~any anisotropy in the MW halo caused by the infall of the Magellanic Clouds may invalidate the assumption of spherical symmetry. Our sample of NGVS RR~Lyrae candidates, with its higher precision single-epoch photometry, multi-year time baseline, and visual vetting to guard against long period variables, is likely to be cleaner of QSOs, even at these large distances.

We compare our measured MW halo density profile slope with predictions from galaxy formation simulations. \citet{2014MNRAS.444..237P} fit the logarithmic slope of the spherically-averaged stellar density profile out to the virial radius for $\sim5000$~MW analogs in the {\it Illustris\/} simulation. Overall, they measured slopes that range from $-5.5 < n < -3.5$ with a mean value of $n\sim-4.5$ by combining the results from light ($6\times10^{11}M_{\odot}<M_{\rm halo}<9\times10^{11}M_{\odot}$) and massive ($9\times10^{11}M_{\odot} < M_{\rm halo}<2\times10^{12}M_{\odot}$) dark matter halos, with more massive halos exhibiting flatter slopes. They also showed that halos that formed more recently (i.e., had a recent merger event) or accreted a larger fraction of their stellar components from satellite galaxies, exhibit flatter stellar halo slopes. Our NGVS slope result, $n = -4.09\pm0.1$, which is slightly flatter than the average simulation slope, may suggest a more massive Galactic halo mass, or a relatively active recent accretion history of the MW's outer halo.

A major shortcoming of our study is the small sky coverage of NGVS (104~deg$^2$) relative to DES \citep[$\sim5,000$~deg$^2$;][]{2021ApJ...911..109S} and PS1 \citep[$\sim30,000$~deg$^2$;][]{2017AJ....153..204S}. The smaller the field of view, the greater the likelihood that the measured MW halo density profile could be biased by underlying substructure. With this caveat in mind, our NGVS MW halo RR~Lyrae sample is consistent with the \cite{2017MNRAS.470.5014S} model predictions for MW analogs: $\sim10$ field RR~Lyrae at $\sim300$~kpc over the whole sky, while we found one such star in 104~deg$^2$. Their Figure~7 shows significant anisotropy in the distribution of distant RR~Lyrae, and their Figure~5 shows substantial variation in the number of such stars across their different halo simulations. 


\subsection{Stellar Populations and the Bailey Diagram}

\begin{figure}
    \centering
    \includegraphics[width=0.45\textwidth]{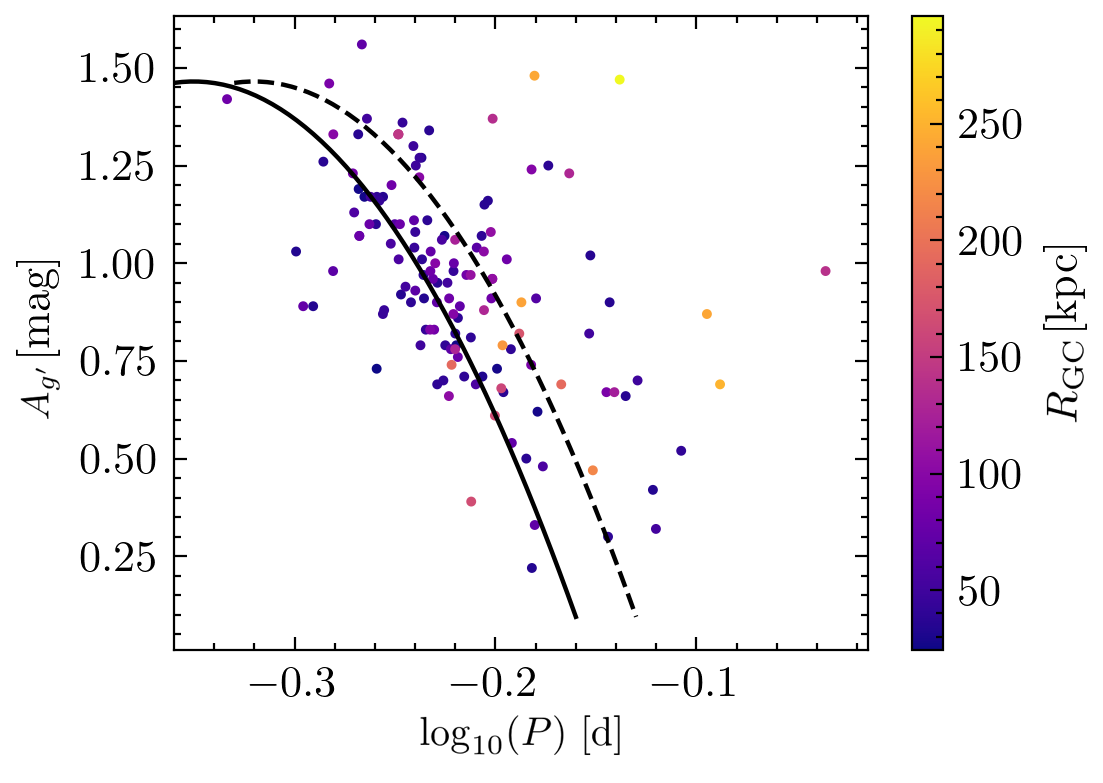}
    \caption{The distribution of our NGVS RR Lyrae on the Bailey diagram, each RR Lyrae is color-coded according to its galactocentric distance $R_{\rm GC}$. RR Lyrae stars within the 100~kpc (black and purple) range concentrates around the Oo~I sequence, while the distant stars (red and yellow) are more sparsely distributed around the Oo~II locus, and have longer periods. The shift from the short-period to the long-period regime indicates the metal-poor environment of outer halo stars' host environments. }
    \label{fig:bailey}
\end{figure}

\citet{1939Obs....62..104O} discovered the bimodality of RRab stars (Oo~I and Oo~II) from different globular clusters (GCs) in period versus amplitude space (the ``Bailey diagram''), and subsequent studies refined this empirical scenario by demonstrating that Oo~I GCs are more metal-rich than Oo~II GCs \citep[e.g.,][]{1959MNRAS.119..134K}. Moreover, the distribution of RR Lyrae on the Bailey diagram is independent of the uncertainties in their distance and reddening. The above factors make the Bailey diagram useful in investigating the formation environment of RR Lyrae. Metallicity differences among RR Lyrae stars from different progenitor dwarf galaxies should be imprinted in the Oosterhoff dichotomy in the Bailey diagram. A recent study by \citet{2019ApJ...882..169F} compiled a catalog of all spectroscopically confirmed field RR Lyrae stars and analyzed their distribution in the Bailey diagram. Unlike RRab stars in GCs, field RRab stars have a continuous (unimodal, rather than bimodal) distribution in the Bailey diagram. While the period, amplitude, and metallicity of field RRab stars do not follow simple linear correlations, RRab stars trend towards being more metal-rich when one moves from long to short periods at fixed amplitude \citep[see Figure 13 of][]{2019ApJ...882..169F}.

\begin{figure}[!t]
    \centering
    \includegraphics[width=0.45\textwidth]{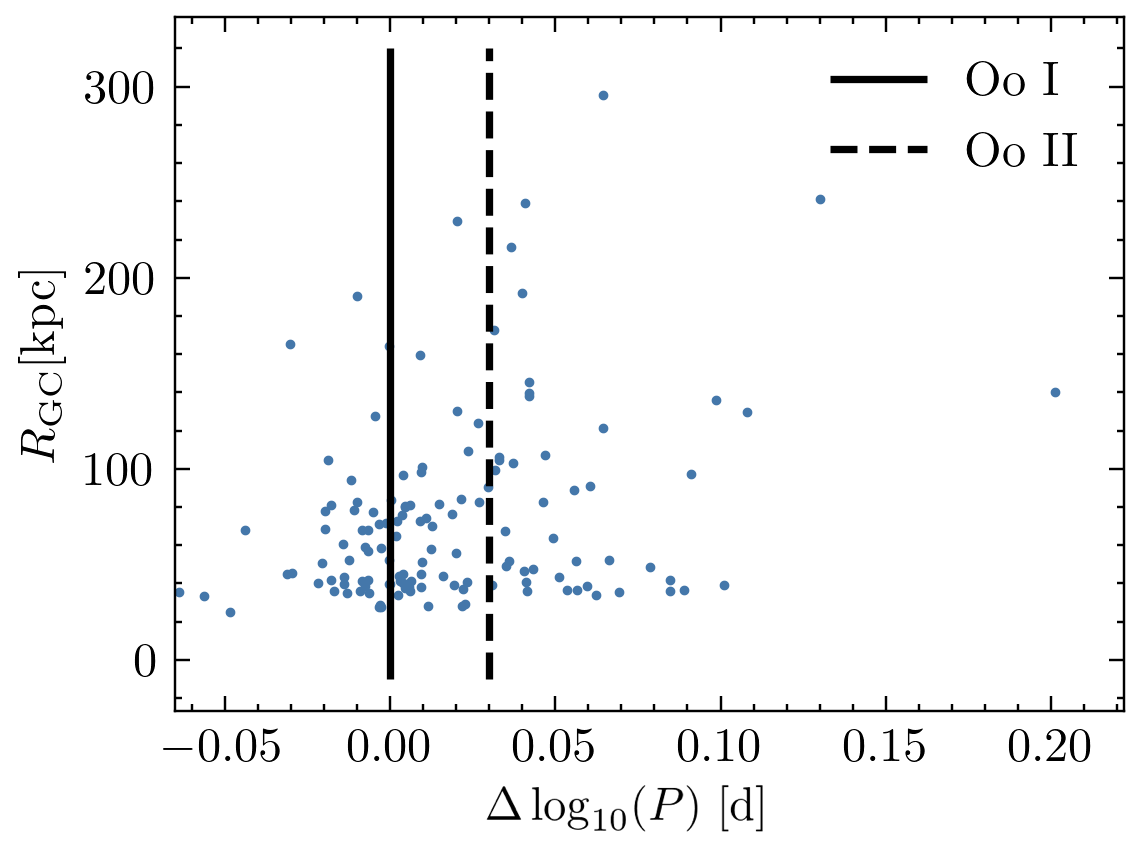}
    \caption{Distribution of each RRab star's x-axis distance towards the Oo~I locus in the Bailey diagram versus $R_{\rm GC}$. We clearly see a cluster of RR Lyrae stars around the Oo~I locus around 100~kpc, while a scattered distribution around the Oo~II locus dominates for stars beyond 200~kpc.}
    \label{fig:dlogp}
\end{figure}
In Figure \ref{fig:bailey}, we show the distribution of our NGVS RRab stars in the Bailey diagram: amplitude $A_{g'}$ (mag) versus the logarithm of the period log$_{10}P$~(d). In Figure \ref{fig:dlogp}, we plot each RRab stars' horizontal distance from the Oo~I locus in the Bailey diagram $\Delta{\rm log}_{10}P$ (d), a rough proxy for metallicity, versus Galactocentric distance $R_{GC}$ (kpc). The analytic relation for the Oo~I locus is a quadratic line fit by \citet{2010ApJ...708..717S}:
\begin{equation}
    \text{Oo~I:}~~~ A_{g'} = -3.18-26.53\log{P} - 37.88(\log{P})^2
\end{equation}
in which $A_{g'}$ is the RRab's $g'$-band amplitude (mag) and $P$ is the pulsation period (d). The Oo~II locus is fitted by offsetting the Oo~I locus by +0.03 in the $\log{P}$ direction. This proxy for the Oo~II locus is the same as the definition in \S\,4.6 of \citet{2010ApJ...708..717S}. The Oo~II line is used to characterize the long-period field RRab subset that do not form a tight secondary sequence. We use the definition: $\Delta \log{P} = \log{P} - \log{P'}$, where $P'$ is the period interpolated from the Oo~I locus at the same $A_{g'}$.

Figure~\ref{fig:dlogp} indicates some degree of clustering around the Oo~I locus for stars within $R_{\rm GC}\sim100$~kpc, but the distribution appears to gradually shift to a broader distribution around the Oo~II locus with increasing $R_{\rm GC}$. This shift in Figure~\ref{fig:dlogp}, or equivalently the systematic shift from the shorter to longer periods in Figure~\ref{fig:bailey}, suggests a change from the metal-rich to the metal-poor regime for the RR~Lyrae stars with increasing Galactocentric distance. Our estimates of the amplitude and period of RR~Lyrae stars may be affected by Blazhko modulations that are undetectable given the sparse sampling of the NGVS RR~Lyrae light curves, especially for distant stars with relatively low-accuracy photometry. This is a possible explanation for the broad distribution of our distant RR~Lyrae around the Oo~II sequence. On the other hand, \citet{2019ApJ...882..169F}, show that the distribution of field RR Lyrae in the Bailey diagram is intrinsically broad and continuous. We believe the variance of the periods of our distant stars reflect the metallicity variance of their progenitor host environments, which are mostly metal-poor for stars beyond 150~kpc. Also, the mean period of our most distant ($R_{\rm GC}>150$~kpc) RRab sample ($P_{\rm mean} = 0.658\pm0.03$~d) is broadly consistent with the recent census results of Gaia RR~Lyrae stars in the nearby ultra faint dwarf (UFD) satellites \citep[$P_{\rm mean} \sim 0.667$~d; see][]{2020ApJS..247...35V}. This, combined with the gradient we see towards longer periods with increasing Galactocentric radius (Figure~\ref{fig:dlogp}), suggests that metal-poor UFD satellites could be a main contributor to the stellar component of the outermost Galactic halo.




\section{Summary and Future Work} \label{sec:future}
We present the detection of 180~RR~Lyrae using observations from the NGVS survey. The data cover 104~deg$^{2}$ of the sky and include in total $\sim38$ epochs across the $u^*$, $g'$, $i'$, and $z'$ bands. The photometric depth of the NGVS data enables us to build a catalog which contains about 100 distant RR~Lyrae that were not included in the PS1 RR~Lyrae search in this region of the sky. We used both light curve template fitting and visual vetting for our RR~Lyrae identification, and we tested the robustness of our RR~Lyrae detection, recovering $96.5\%$ of known, unsaturated RR~Lyrae in the PS1 catalog, with a period match at the $0.2\%$ difference level. Most of the additional RR~Lyrae we contribute have $\langle g'\rangle > 20.5$~mag, corresponding to $d_{\rm helio} > 100$~kpc. The depth of our photometry, multi-year time baseline, and our visual vetting for long-term variability all contribute to a low QSO contamination rate for our distant RR~Lyrae sample.

We fit a spherical stellar halo model to the radial number density profile of our RR Lyrae beyond 45 kpc (Figure \ref{fig:density}), and obtained a single power law slope of $n = -4.09\pm0.10$, out to a distance of $R_{\rm GC} \sim 300$~kpc. The existence of RR~Lyrae out to 300~kpc without any clear density break is consistent with the result of the DES Y6 RR~Lyrae catalog, and also supports the theoretical prediction of the splash back radius of MW halo being at least at this distance \citep[$\sim\!0.8 r_{200\rm m} = 290\pm 61$~kpc,][]{2020MNRAS.496.3929D}. The radial density slope of stellar halo has been studied by previous works using various tracers including RR Lyrae, K giants and BHB stars, with estimated slopes between $n = -3.8\pm0.1$ to $n=-5.4\pm0.1$. Our result is broadly consistent with these literature values, yet on the flatter end. By comparing our slope with halo simulation results in \citet{2014MNRAS.444..237P}, which fit power-law stellar density profiles for more than 5000 MW mass halos, our estimated slope supports a more massive Galactic halo mass, or a relatively active recent accretion history of the outer halo. 

We examine the distribution of our NGVS RRab stars in the Bailey diagram of period $P$ versus amplitude $A_{g'}$, and find a clear evidence of an Oosterhoff dichotomy between inner halo and outer halo RRab stars (Figure \ref{fig:bailey} and Figure \ref{fig:dlogp}), where 
a substantial group of our NGVS RRab stars within 100~kpc cluster around the Oo~I (more metal-rich) locus, and as distance grows our outer halo RRab stars gradually shift to the longer-period regime and follow a scattered distribution around the Oo~II (more metal-poor) locus. For outer halo stars with $R_{\rm GC} > 200$~kpc, nearly no RRab is found around the Oo~I locus. Our result supports the idea that metal-poor UFD satellites could be the main contributor of stars to the outermost Galactic halo. 

Our future work on our NGVS RR~Lyrae sample will be in two general directions. First, we plan to work on a statistical characterization of the level of contamination by QSOs and contact binaries. We plan to model the brightness variation behaviour of these objects in sparsely sampled surveys like NGVS by subsampling known QSOs and contact binaries in the Canada-France-Hawaii Telescope Legacy Survey (CFHTLS) Deep Fields data base which has a 10-yr observation baseline and more than 1,500 measurements across the $u^*$, $g'$, $r'$, $i'$, and $z'$ bands. By creating mock QSO and binary light curves and and processing them through our RR~Lyrae identification machinery, we expect to derive empirically calibrated estimates of purity as a function of distance for our NGVS RR~Lyrae sample. Second, we are in the process of obtaining medium resolution spectra for our newly identified RR~Lyrae, and analyzing their kinematical distributions. 

The results and methodology presented in this work may  be helpful for future research on RR~Lyrae stars in the coming age of Vera C.\ Rubin Observatory \citep{2009arXiv0912.0201L}. Rubin data will provide a huge leap in capability for finding variable objects in the Galactic halo, with its unparalleled temporal coverage and photometric depth. Our work, together with the PS1, HiTS, and DES results, will serve as good training sets and pathfinders to help calibrate the detection of RR Lyrae objects in Rubin data.

\acknowledgments
We acknowledge the helpful comments from our anonymous referee, which improved the paper significantly. YF and PG acknowledge support from the National Science Foundation grant AST-2206328. This paper is based on observations obtained with MegaPrime/MegaCam, a joint project of CFHT and CEA/IRFU, at the Canada-France-Hawaii Telescope (CFHT), which is operated by the National Research Council (NRC) of Canada, the Institut National des Sciences de l’Univers of the Centre National de la Recherche Scientifique (CNRS) of France, and the University of Hawaii. This research used the facilities of the Canadian Astronomy Data Centre operated by the National Research Council of Canada with the support of the Canadian Space Agency. This research used the Canadian Advanced Network For Astronomy Research (CANFAR) operated in partnership by the Canadian Astronomy Data Centre and The Digital Research Alliance of Canada with support from the National Research Council of Canada the Canadian Space Agency, CANARIE and the Canadian Foundation for Innovation. We thank the following University of California Santa Cruz undergraduate students for their help with the visual vetting: Kayla Bartel, Spencer Jaseph, Kyle Nguyen, Talise Oh, and Casey Peters. We also express our gratitude to a group of high school members of the research {\bf team} who were interns in the Science Internship Program (SIP) at UCSC.

\facilities{CFHT}
\software{Astropy \citep{astropy:2013, astropy:2018, astropy:2022}}




\bibliography{AAS46410}{}
\bibliographystyle{aasjournal}



\end{document}